\documentclass[sigconf,screen,prologue,table,xcdraw]{acmart}

\acmConference[ISSTA '23]{Proceedings of the 32nd ACM SIGSOFT International Symposium on Software Testing and Analysis}{July 17--21, 2023}{Seattle, WA, USA}
\setcopyright{acmcopyright}
\copyrightyear{2023}
\acmYear{2023}
\acmPrice{15.00}

\usepackage{hyperref}[unicode]
\usepackage{multirow}
\usepackage{graphicx}
\usepackage{tabularx}
\usepackage[frozencache=true,cachedir=minted-cache]{minted}
\usepackage{url}

\newcommand{\emulatorName}[0]{{\sc Icicle}}
\newcommand{\emulatorNameBold}[0]{{\sc \textbf{Icicle}}}
\usepackage[framemethod=TikZ]{mdframed}

\newcommand{\benchmarkCPU}[0]{AMD Ryzen Threadripper 3990X}
\newcommand{\aflVersion}[0]{{AFL++ 4.01a}}

\newcommand{\iaArch}[0]{{x86}}

\usepackage{enumitem}
\setlist[itemize]{align=parleft,left=1em..2em}

\usepackage{tikz}
\usepackage{amsmath}
\usepackage{multirow}

\usepackage[utf8]{inputenc}
\DeclareUnicodeCharacter{2713}{{\checkmark}} 
\DeclareUnicodeCharacter{2717}{{-}} 

\newcommand{\circled}[2][]{
  \tikz[baseline=(char.base)]{%
    \node[anchor=text, shape=circle,draw, inner sep=0pt, minimum size=0.5em] (char){#1\strut};
    \node at (char.center) {\makebox[0pt][c]{#2}};}}
\robustify{\circled}

\usepackage{array}
\usepackage{dblfloatfix} 

\begin{document}

\title{\emulatorName{:} A Re-designed Emulator for Grey-Box Firmware Fuzzing}

\author{Michael Chesser}
\affiliation{%
      \institution{The University of Adelaide\\
      Data61 CSIRO, Cyber Security Cooperative Research Centre}
      \country{Australia}
}
\email{michael.chesser@adelaide.edu.au}

\author{Surya Nepal}
\affiliation{%
      \institution{Data61 CSIRO, Cyber Security Cooperative Research Centre}
      \country{Australia}
}
\email{Surya.Nepal@data61.csiro.au}

\author{Damith C. Ranasinghe}
\affiliation{%
      \institution{The University of Adelaide\\
      The School of Computer Science}
      \country{Australia}
}
\email{damith.ranasinghe@adelaide.edu.au}

\begin{abstract}
Emulation-based fuzzers enable testing binaries without source code and facilitate testing embedded applications where automated execution on the target hardware architecture is difficult and slow. The instrumentation techniques added to extract feedback and guide input mutations towards generating effective test cases is at the core of modern fuzzers. But, modern emulation-based fuzzers have evolved by re-purposing general-purpose emulators; consequently, developing and integrating fuzzing techniques, such as instrumentation methods, is difficult and often added in an ad-hoc manner, specific to an instruction set architecture (ISA). This limits state-of-the-art fuzzing techniques to a few ISAs such as x86/x86-64 or ARM/AArch64; a significant problem for \textit{firmware fuzzing} of diverse ISAs.

This study presents our efforts to \textit{re-think emulation for fuzzing}. We design and implement a fuzzing-specific, multi-architecture emulation framework---\emulatorName{}. We demonstrate the capability to add instrumentation once, in an \textit{architecture agnostic} manner, with low execution overhead.
We employ \emulatorName{} as the emulator for a state-of-the-art ARM firmware fuzzer---{\sc Fuzzware}---and replicate results. Significantly, we demonstrate the availability of new instrumentation in \emulatorName{} enabled the discovery of new bugs. We demonstrate the fidelity of \emulatorName{} and efficacy of architecture agnostic instrumentation by discovering bugs in benchmarks that require a known and \textit{specific} operational capability of instrumentation techniques, \textit{across a diverse set of instruction set architectures} (\iaArch{}, ARM/AArch64, RISC-V, MIPS). Further, to demonstrate the effectiveness of \emulatorName{} to discover bugs in a currently \textit{unsupported} architecture in emulation-based fuzzers, we perform a fuzzing campaign with real-world firmware binaries for Texas Instruments' MSP430 ISA and discovered 7 new bugs.

\end{abstract}



\maketitle

\section{Introduction}\label{introduction}

Fuzzing is an automated software testing methodology that repeatedly executes a program with generated inputs and monitors execution for adverse behaviors. Progress in the field has greatly enhanced the \textit{bug discovery} capability of modern fuzzers and fuzzing is now widely used in the software development industry. In particular, grey-box fuzzing (or feedback-driven) methods have proven to be highly effective at scale and are capable of finding bugs in a diverse set of software~\cite{grimoire2019, nautilus2019, parmesan2020, nyx2020, magma2020, fuzzbench2021, snapfuzz2022, printfuzz2022}. Grey-box fuzzing relies on the ability to add instrumentation to the target program to obtain feedback. This feedback allows input generation to be intelligently guided, improving a fuzzer's ability to discover bugs.

A simple method to facilitate grey-box fuzzing is for the compiler to inject instrumentation into the source code during compilation. However, it is often necessary to fuzz binaries where source code is unavailable---\textit{binary-only fuzzing}---or where the target hardware is not suitable for automating testing and testing is carried out on a host machine with a different instruction set architecture (ISA)---\textit{cross-architecture fuzzing}. For instance, it is extremely challenging to perform rapid execution on devices typically used for Internet of Things (IoT) applications and embedded systems in general~\cite{firmadyne2016, embedded2016, avatar2018, pretender2019, unicorefuzz2019, halucinator2020, firmcorn2020, p2im2020, equafl2022}. Consequently, we are forced to use \textit{emulators} capable of executing binaries built for the target on a more convenient host machine; exploiting the resource capabilities of the host for software bug discovery and triaging. Therefore, emulators play a critical role in supporting both \textit{binary-only} and \textit{cross-architecture} fuzzing. Significantly, emulators enable unparalleled control and introspection over program execution, even without source code and access to the original hardware.

Current state-of-practice for emulation-based grey-box fuzzing, driven by its more recent evolution compared to emulators, is to integrate fuzzing instrumentation into existing general-purpose emulators. But, this can be challenging because these emulators were not designed to support such modifications~\cite{symqemu2021}. Consequently, existing emulation-based fuzzers implement instrumentation techniques either manually, through direct modifications to the emulator~\cite{afl2010, aflplusplus2020, qasan2020, symqemu2021}, or through limited interfaces that are unable to support more advanced instrumentation~\cite{unicorn2015, unicorefuzz2019, firmcorn2020, basesafe2020}. Further, the absence of a consistent approach to add new, experimental, instrumentation and the need for domain expertise in emulator development to evaluate new fuzzing techniques are arguably barriers to developments in the field. As a result, the benefits of extensive research efforts to develop state-of-the-art fuzzing techniques can remain limited to a specific ISA; this is \textit{undesirable}.

\vspace{2mm}
\noindent\textbf{Our Contributions.~}This study presents our efforts to design and build a \textit{new} multi-architecture emulation framework explicitly for fuzzing.

In summary, we make the following contributions:

\begin{itemize}
\item
    We designed a new multi-architecture emulator, \emulatorName{}, for directly supporting emulation-based fuzzing: i)~enabling the implementation of architecture-agnostic instrumentation; ii)~employing a decoupled design, enabling emulation, instrumentation and instruction set architecture (ISA) support to be maintained separately; and iii)~byte level memory-management to better support emulating memory in embedded systems.

\item
    We implemented \emulatorName{} as a coverage-guided greybox fuzzer by integrating with AFL++ and {\sc Fuzzware}.

\item
    We conducted extensive experiments across five diverse ISAs, multiple instrumentation techniques, 21 real-world binaries and a synthetic test program. Notably, we \textit{demonstrate}: i)~the instrumentation requirements of state-of-the-art fuzzing techniques are satisfied with a unified instrumentation interface without the need for architecture-specific knowledge; ii)~the effectiveness of \emulatorName{} by comparing against existing emulators for the challenging task of firmware fuzzing. Significantly, \emulatorName{}, supported by additional architecture-agnostic instrumentation, uncovered previously undiscovered bugs; and iii)~the efficacy of \emulatorName{} and its architecture agnostic instrumentation on a new ISA---we fuzz and discover seven bugs in real-world binaries written for Texas Instruments' MSP430 RISC architecture. Importantly, this ISA is currently \textit{not supported} by existing emulation-based fuzzers.

\item
    We \textit{\textbf{open source}}\footnote{\url{https://github.com/icicle-emu/icicle}} \emulatorName{} to facilitate further improvements and advance the field of emulation-based fuzzing in general.
\end{itemize}

\section{Instrumentation Challenges in Emulation-based Fuzzing}\label{background}

In this section, we present an overview of different instrumentation techniques that are used in modern grey-box fuzzing frameworks. Subsequently, we highlight the issues hindering their implementation in existing general-purpose emulators without resorting to direct, architecture-specific, modifications of the emulator that motivate the need of a re-designed emulator for fuzzing.

\subsection{Instrumentation Techniques}\label{sec:instrumentation-overview}

Grey-box fuzzers rely on instrumentation techniques to obtain feedback to enable more effective exploration of a target program which is necessary for uncovering deeper bugs. Therefore, it is important for fuzzing frameworks to support a diverse set of instrumentation techniques. In this section we describe several instrumentation techniques developed in previous research efforts.

\vspace{1mm}
\noindent\textbf{Code coverage (Branch hit counts).}~Almost all grey-box fuzzers utilize a form of code coverage for feedback. Code coverage identifies inputs that reach new locations within a program by instrumenting the target so it notifies the fuzzer when new code is reached. It is a proven and effective method that enables fuzzer to incrementally discover different parts of the program~\cite{fuzzbench2021, covbenchreliability2022}. Branch hit counts is an approach to code coverage popularized by AFL~\cite{afl2010}. This approach maintains a global map of 8-bit counters that are incremented whenever an edge in the program is hit. After execution, the values of each of the counters grouped into one of 8 ranges and if any of the counters contain a value with a unique range, the fuzz input corresponding to the execution is considered novel.

\vspace{1mm}
\noindent\textbf{Context-sensitive branch coverage}~\cite{angora2018, aflsensitive2019}.~Context-sensitive branch coverage augments branch hit counts by hashing the edge index with the current calling context. This allows the fuzzer to obtain better feedback from branches inside of frequently called functions.

\vspace{1mm}
\noindent\textbf{CmpLog}~\cite{aflplusplus2020}.~For many target binaries, code coverage is insufficient at finding inputs that reach deep parts of the program. For example, comparisons against large constants (such as Listing \ref{hard-comparison}) are difficult to satisfy with code coverage alone, because there is no feedback mechanism that allows for incremental progress towards solving the comparisons.

\begin{listing}[htp]
\begin{minted}{c}
  if (tag == 0x31677562) {
    crash();
  }
\end{minted}
\caption{Example program where \texttt{crash} is hard to reach with traditional code coverage instrumentation.}
\label{hard-comparison}
\end{listing}

One approach explored in several recent studies~\cite{redqueen2019, greyone2020, weizz2020} is to directly instrument the operands of comparisons. CmpLog is a comparison tracing technique implemented in AFL++ based on {\sc Redqueen }~\cite{redqueen2019} and {\sc Weizz }~\cite{weizz2020}. CmpLog identifies comparisons within a program, then adds instrumentation that captures the operands of the comparison. After execution, the fuzzing frontend can then scan the input for the operands in order to replace them with their correct value (a process referred to as input-to-state replacement).

\vspace{1mm}
\noindent\textbf{CompareCov}~\cite{aflplusplus2020}.~Is an alternative approach to solving complex comparisons based an earlier compiler instrumentation technique, LAF-Intel~\cite{lafintel2016}. CompareCov provides better feedback by instrumenting the program to split comparisons involving large values into comparison between individual bytes. The split comparison can subsequently be solved with code coverage instrumentation.
In addition to instruction-level comparisons, CompareCov also attempts to improve feedback for memory comparison functions (\texttt{memcmp}, \texttt{strcmp} and \texttt{strncmp}) by adding instrumentation to update the coverage (bitmap) for every matching byte within the comparison operation.

\begin{figure*}[b]
\centering
\includegraphics[width=0.95\linewidth]{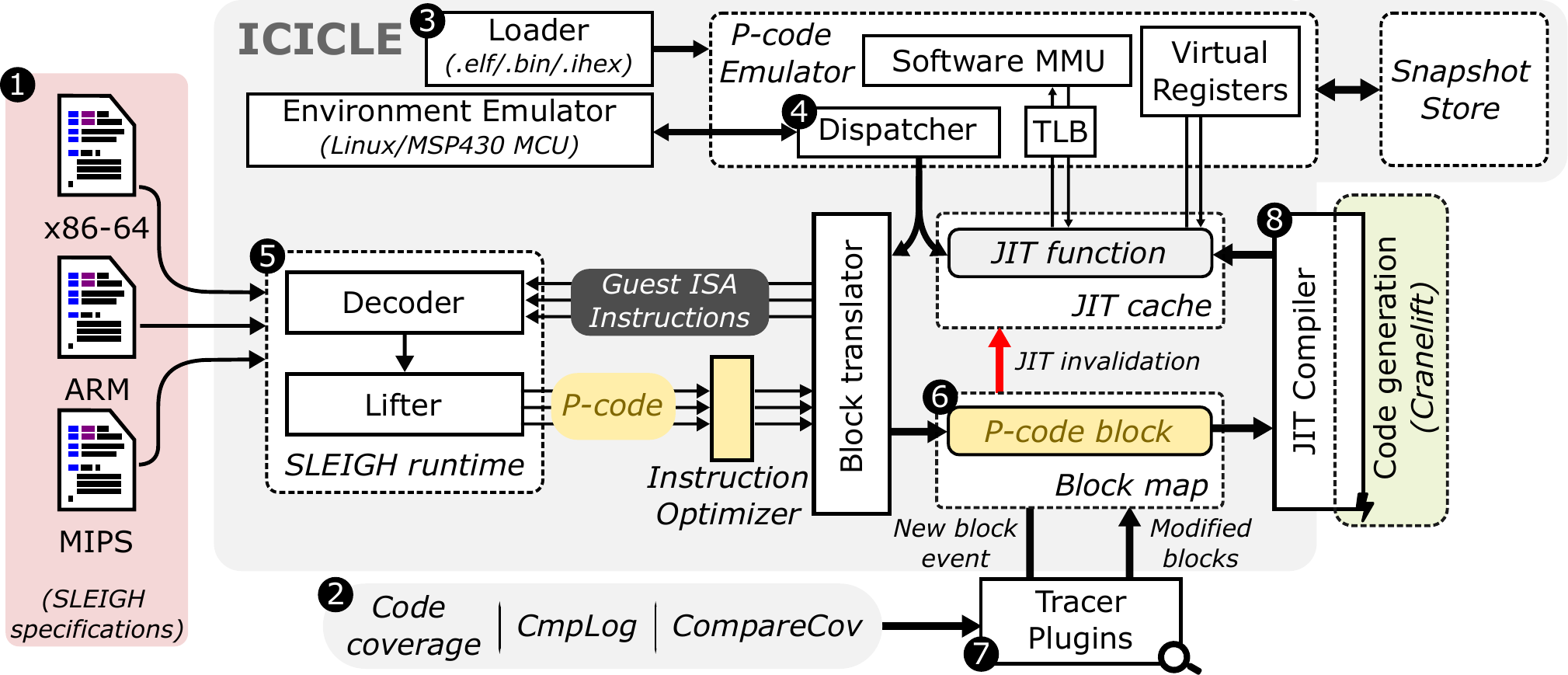}
\caption{Overview of the core components in our fuzzing specific multi-architecture emulation framework. The instrumentation workflow is as follows:
\circled{1} On initialization the emulator loads the appropriate SLEIGH processor specification, configuring the SLEIGH runtime. \circled{2} One or more \textit{Tracer Plugins} are registered with the emulator to support the instrumentation needs of the fuzzer. \circled{3} Once configured, the emulator loads the target binary into memory and starts execution. \circled{4} During execution, whenever the emulator attempts to execute a new instruction, the dispatcher initiates the translation process. \circled{5} Each guest instruction is then translated to P-code. \circled{6} P-code operations are grouped into a block, and the block is stored in a global block map. \circled{7} Tracer Plugins are notified to allow them to analyze the new block and modify any block in the block map required for instrumentation. \circled{8} Any new or modified blocks are then compiled to native code using the JIT compiler and the emulator resumes execution. Notably, \textit{all} of components within the grey area were implemented as part of \emulatorName{} and only the SLEIGH specifications we employed were from Ghidra.} 
\label{fig:emulator-overview}
\end{figure*}

\subsection{Instrumentation Challenges}

QEMU and Unicorn (forked from QEMU), have become the de facto emulators for fuzzers~\cite{bitblaze2008, s2e2011, triforceafl2016, avatar2018, pretender2019, firmafl2019, unicorefuzz2019, firmcorn2020, halucinator2020, inception2018, p2im2020, basesafe2020, symqemu2021, equafl2022, uemu2021, fuzzware2022}. Therefore, for the remainder of this paper we will primarily compare \emulatorName{} against emulator-level capabilities in QEMU and Unicorn to support fuzzing.

QEMU was primarily designed for fast, general-purpose, emulation, not for fuzzing. Consequently, many design decisions differ from those important for fuzzing. Crucially, it is difficult to add advanced instrumentation techniques in an architecture agnostic manner, for the following key reasons:

\begin{itemize}
\item
      QEMU's intermediate representation for dynamic translation, Tiny Code Generator (TCG) ops, is not designed for direct analysis or manipulation~\cite{symqemu2021}. Consequently, it is challenging to add instrumentation without making invasive code changes at a very low level.

\item
      QEMU is monolithic in design, providing little support for extensibility, and assumes that it controls the full life cycle of the emulation process. These properties inhibit the maintainability of modifications and reduce scalability as it makes it challenging to share state across fuzzing instances.

\item
      Given QEMU's historical focus on emulation, there is no unified mechanism for adding instrumentation. For example, code coverage is implemented by modifying the code-generator to inject code at the start of every basic block; more complex techniques, such as CmpLog, modify the translation process of individual architectures. Hence, adding new instrumentation techniques require extensive emulation domain expertise.
\end{itemize}

Unicorn is a fork of QEMU designed for more flexible and modular CPU emulation. Unicorn extracts the CPU emulation component from QEMU, configures it to always use the software MMU implementation, and introduces hooks, functions that are called in response to selected emulator events, such as memory accesses and breakpoints~\cite{unicorefuzz2019, basesafe2020, fuzzware2022}.

Unicorn's function hooking API enables fuzzers to inject functions calls to observe the CPU state, which can be used to implement some instrumentation techniques. However, the observable state is architecture specific, and Unicorn provides no support for analyzing the code semantics, making more advanced instrumentation difficult. Additionally, the maintainers of the Unicorn project have also reported that it is difficult to keep Unicorn up-to-date with improvements because of QEMU's monolithic design~\cite{unicorn2github2020}.

Therefore, we are motivated to build a new multi-architecture emulation framework explicitly designed for fuzzing; with the ability to support sophisticated instrumentation methods in an architecture-agnostic manner to enable fast emulation-based fuzzing of binaries.

\section{Icicle Design and Implementation}\label{implementation}

We provide an overview of our fuzzing specific multi-architecture emulation framework in Figure~\ref{fig:emulator-overview}. To enable architecture agnostic fuzzing we use a portable intermediate representation (IR) that is suitable for both \textit{emulation and program analysis}. Translation of the guest ISA to the portable IR is achieved using processor specifications that are external to the emulator. This ensures that architecture-specific details are kept decoupled; enabling fixes for specification bugs and the addition of new architectures to be implemented with minimal changes to the core emulator. Further, in contrast to existing emulation-based fuzzing frameworks, we define new instrumentation application programming interfaces (APIs) that enable instrumentation to exist entirely outside of the emulator. This facilitates researchers both, in developing new instrumentation techniques for emulation-based fuzzers without domain expertise in emulator development, and \textit{the immediate availability of these techniques across ISAs} in a design-build-and-test \textit{once-only} paradigm.

\subsection{Fuzzing Specific Emulator Design}

\emulatorName{} supports fast, multi-architecture, CPU emulation through portable dynamic translation. First, guest ISA instructions are translated to an intermediate representation (IR) called P-code. P-code is then just-in-time (JIT) compiled to the host architecture allowing for efficient execution.

\emulatorName{} performs translation to P-code through the use of SLEIGH\footnote{The name SLEIGH was derived from SLED (Specification Language for Encoding and Decoding)\cite{ghidra2019}, which also influenced the name of our emulator: \emulatorName{}} processor specifications for the guest ISA. SLEIGH is a domain-specific language (DSL) that describes how to decode and translate the semantics of machine code into P-code. We chose SLEIGH as the basis of \emulatorName{'s} CPU emulation for the following reasons:

\begin{itemize}
\item
    \textit{Broad architecture support}. We leverage the diverse set of SLEIGH processor specifications that have already been created as part of the open-source Ghidra framework~\cite{ghidra2019} (over 45 processor kinds are supported). This enables \emulatorName{} to emulate a wide range of architectures, including architectures \textit{unsupported} by other emulation frameworks like QEMU, with significantly reduced effort.

\item
    \textit{Designed for analysis}. Unlike IRs used in other emulator frameworks, P-code was explicitly designed to support program analysis. For example, P-code maintains hints for call and return operations even though such hints are unnecessary for emulation. This makes it better suited for performing the code analysis tasks required for advanced instrumentation techniques.

\item
    \textit{Suitable for emulation}. P-code consists of small set of operations that can be efficiently executed by the host ISA. For example, P-code avoids the use of bit-vectors. This allows for fast emulation.

\item
    \textit{Decoupling and ease of maintenance}. \emulatorName{} uses the original SLEIGH specifications written for Ghidra without any modifications. This allows any improvements or fixes made to a specification to be immediately usable by \emulatorName{} without any changes to the core emulator. Importantly, the use of SLEIGH specification naturally enables \emulatorName{} to emulate new targets by \textit{simply} providing the new SLEIGH specification. Significantly, the new architecture will benefit from any existing instrumentation techniques \textit{without} needed change to the emulator.
\end{itemize}

\emulatorName{} implements a P-code emulator\footnote{Ghidra contains a limited P-code emulator and has been used for micro-fuzzing~\cite{aflghidraemu2021} but is unable to satisfy the needs of a full modern fuzzing framework, for example, Ghidra's P-code emulator is interpreter-based hindering performance. Notably, \emulatorName{} only uses the SLEIGH specifications and \textit{none} of the components of Ghidra's emulator.} consisting of a SLEIGH runtime (a SLEIGH processor specification compiler, decoder and lifter), a JIT-based execution engine and an efficient software memory-management unit (MMU) implementation. The SLEIGH runtime handles loading the appropriate SLEIGH specification for the guest architecture, assigning a mapping from guest registers to virtual P-code registers, and then decoding and translating ISA-specific machine code to P-code. Unlike Ghidra's SLEIGH runtime, \emulatorName{'s} runtime assigns sequential IDs to virtual registers, allowing them to be managed in a dense array, improving performance. We also implement a lightweight P-code optimization pass that performs constant evaluation and dead-code elimination, significantly reducing the number of P-code operations when values are known at translation time.
\emulatorName{}'s JIT-based execution engine, then groups P-code operations in blocks and compiles them to native code using Cranelift~\cite{craneliftgithub2021}, an open-source low-level code generation framework. Cranelift provides register allocation, instruction legalization, and additional optimizations. Later, during a recompilation step, multiple blocks are compiled as part of a single compilation unit, enabling additional optimizations that improve performance. Notably, \textit{unlike} existing emulators, \emulatorName{} does not discard the P-code representation of each block after JIT compilation. This can significantly aid any analysis used for complex instrumentation, at the cost of some additional memory overhead.

\begin{figure}[!htbp]
\centering
\includegraphics[width=0.7\linewidth]{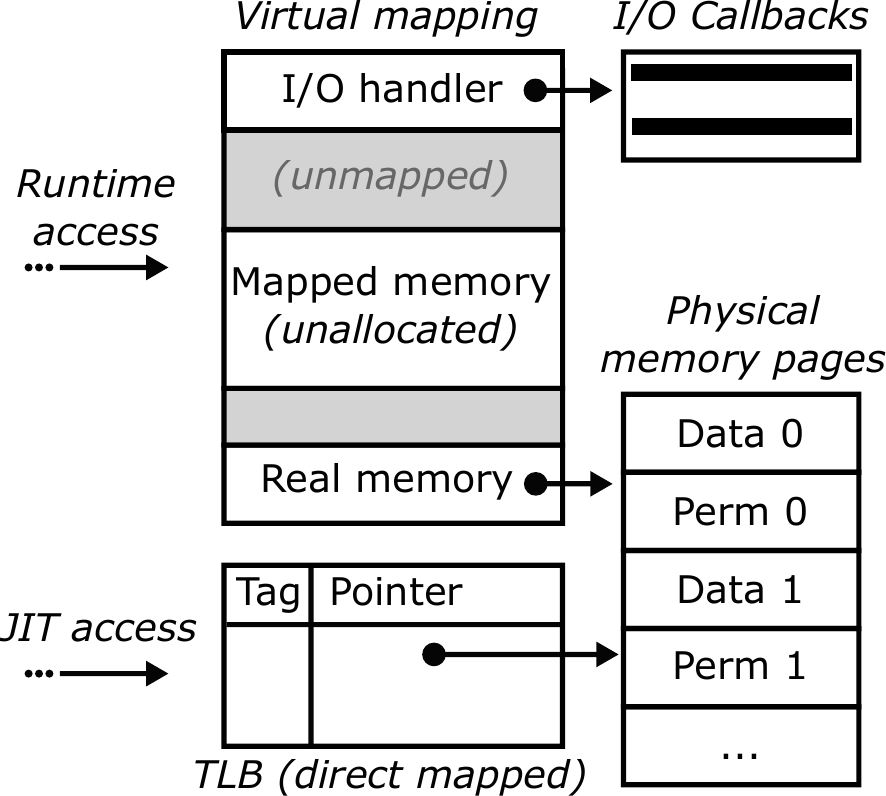}
\caption{Overview of the byte-level software memory-management unit (MMU) implemented in \emulatorName{}.}
\label{fig:softmmu}
\end{figure}

When the memory layout of the guest is incompatible with the host, it is necessary for the emulator to handle the differences. Therefore, \emulatorName{} uses a software MMU to handle guest memory accesses (Figure~\ref{fig:softmmu}). The software MMU maintains virtual mapping table that allows guest memory and memory mapped IO (MMIO) to be mapped in the emulator. The mapping table is represented as a range-map, allowing for \textit{byte}-level precision at the cost of more expensive lookups. To improve efficiency, we cache translated addresses in a lookup table referred to as a translation lookaside buffer (TLB), named after its hardware analog. The JIT compiled code can directly access guest memory using the TLB allowing for fast execution in most cases. Whenever, an address not in the TLB, is encountered, the JIT calls a runtime helper that handles the access and caches the translated address. To retain byte-level mapping, \emulatorName{} maintains a permission byte for each physical byte which is checked by the JIT on access, similar to approaches used in prior work~\cite{softserve2018, qasan2020, sfuzz2022}.  Both Unicorn and QEMU (when running in full-system mode) also implement a software MMU. However, they both require memory to be mapped in page-sized (4~KB) regions.

\textit{The added flexibility of the byte-level mapping in \emulatorName{} allows more accurate emulation of embedded system memory and can be used to enable better bug detection}.

In addition to CPU emulation, most binaries interact with external resources such as file systems, hardware, and other software. \emulatorName{} is designed to be flexible and extensible using pluggable \textit{environment emulators} (see Figure~\ref{fig:emulator-overview}). To demonstrate the functionality of the system, we have implemented environments to allow comparisons against existing emulation-based fuzzers. The current implementation supports fuzzing Linux userspace binaries by emulating a subset of system calls, supports fuzzing several MSP430 MCUs ISAs, and supports embedded ARM binaries using {\sc Fuzzware}~\cite{fuzzware2022}.

\subsection{Architecture-Agnostic Instrumentation}\label{instrumentation}

To implement arbitrarily complex fuzzing instrumentation requires: i)~the ability to analyse the semantics of a target program; ii)~an efficient mechanism to capture runtime information about the running program;  iii)~a way of sharing the captured information with the fuzzing frontend. Additionally, to be effective in a fuzzing context, these requirements must be supported in a manner that has low performance overheads.

\emulatorName{} supports these requirements through a set of APIs added to the emulator, we refer to instrumentation utilizing these APIs as \textit{Tracer Plugins}. These APIs enable:
\begin{itemize}
\item \textit{Direct access to the architecture-agnostic P-code representation of the program}. Plugins are able register a callback function to be called whenever the emulator translates a new block to P-code. The callback function is provided with the full code-cache including the newly translated block, satisfying the first requirement enabling architecture agnostic code analysis.
\item \textit{Inline code-injection}. Plugins can inject additional P-code operations into any block enabling inline instrumentation to be supported in an architecture agnostic manner. Any modified blocks are invalided by \emulatorName{} and re-compiled by the JIT the next time they are executed.
\item \textit{Registry of JIT and fuzzer accessible shared memory}. During initialization, plugins can register storage locations with the emulator, which can later be manipulated with P-code operations. Additionally, \emulatorName{} allows plugins to define custom P-code registers, these custom registers are treated the same as guest registers for the purpose of register allocation during JIT code-generation, which can allow for more efficient instrumentation in some cases. This enables data to be efficiently saved by injected instrumentation and analyzed as part of the fuzzing loop.
\end{itemize}

To illustrate expressiveness \emulatorName{'s} instrumentation method and its ability to support architecture-agnostic instrumentation, we discuss the implementation of the four techniques discussed in Section.~\ref{sec:instrumentation-overview} in \emulatorName{} and compare them to implementations in other emulation-based fuzzers.

\vspace{1mm}
\noindent\textbf{Branch hit counts.~}In \emulatorName{}, branch hit counts are implemented by a Tracer Plugin that does the following: during initialization, it registers the location of coverage bitmap with the emulator and defines a custom register to store the previous program location. When a new block is translated, the plugin injects code at the start of the block that computes a hash of (current\_location, previous\_location), which is then used as an index for updating the coverage bitmap. Since the instrumentation is implemented using P-code injections, the JIT can generate native code that updates the coverage bitmap without resorting to a function call. In addition to branch hit counts, \emulatorName{} also implements both block-only coverage and edge coverage.

Existing emulators are also able to add branch hit count instrumentation in an architecture independent manner by injecting code when new translation blocks are created, which is common across architectures in QEMU. However, AFL++'s implementation in both QEMU and Unicorn makes direct modifications the emulator (although the changes are relatively minor). Notably, in AFL++'s Unicorn mode, branch coverage instrumentation is \textit{not} implemented using Unicorn's hooking API, since the instrumentation is highly performance sensitive, and the hooking API imposes additional overheads.

\vspace{1mm}
\noindent\textbf{Context-sensitive branch coverage.~}\emulatorName{} implements context-sensitive branch coverage with a Tracer Plugin. This plugin defines a custom register to store the context, then when a new block ending with a \texttt{CALL} is translated, the generates a random value to use as context for the current location and XORs it the context register. In the block after the call, the instrumentation is injected to clear the added context. The branch hit count plugin is then modified to use the context value by using it as part of computing the index into the coverage map. Since the \texttt{CALL} hint is part of P-code representation, it allows us to write a portable implementation that works across architectures.

Context-sensitive coverage was first implemented in Angora\cite{angora2018} using compiler-based instrumentation, and in afl-sensitive~\cite{aflsensitive2019} for binary-only instrumentation using QEMU. afl-sensitive's implementation modifies QEMU's \iaArch{} translation layer to add instrumentation that updates the calling context on \texttt{call} and \texttt{ret} instructions. Since afl-sensitive instruments \iaArch{} specific instructions it is not portable to other architectures.

\vspace{1mm}
\noindent\textbf{CmpLog.~}There are two parts to CmpLog, first relevant comparison operations must be identified, and second, the operands of each comparison must be copied to a fuzzer accessible location.

Inspired by the success of Datalog for program analysis tasks~\cite{datalogdisasm2020, usingdatalog2010}, we implement a comparison finding algorithm as a set of Datalog rules in Listing~\ref{listing:compare}. Since the rules are defined in terms of P-code operations, it allows \emulatorName{} support CmpLog for any architecture.

\begin{listing}[!htbp]
\begin{minted}[fontsize=\footnotesize,framesep=2mm]{prolog}
% x is an copy of the destination of an operation.
copy(x, x) :- op(x, _, _, _).
% b = a if it is the destination of a copy-like op with a.
copy(a, b) :- op(b, "COPY", a).
copy(a, b) :- op(b, "ZXT", a).
% b = a if x = a and b = x
copy(a, b) :- copy(a, x), copy(x, b).
% Identify p-code operations corresponding to comparisons.
cmp("==", cond, a, b) :- op(cond, "==", a, b).
cmp("!=", cond, a, b) :- op(cond, "!=", a, b).
% `(a - b) [cmp] 0` => `a [cmp] b` (subtract and compare with zero)
cmp(op, cond, a, b) :- op(cond, "-", a, b), cmp(op, cond, x, 0).
% `!(a [inv(op)] b)` => `a [op] b` (inverted comparison)
cmp("!=", cond, x, y) :- op(notc, "!", cond), cmp("==", notc, x, y).
cmp("==", cond, x, y) :- op(notc, "!", cond), cmp("!=", notc, x, y).
% Output comparisons that flow into the branch condition.
output(op, a, b) :- cmp(op, cond, a, b), copy(cond, x), branch(x).
\end{minted}
\caption{Datalog rules for finding comparison operands. The list of p-code operations to analyse, and the branch exit condition are provided as inputs.}
\label{listing:compare}
\end{listing}

In contrast, existing CmpLog implementations require identifying architecture specific instructions to identify comparisons. For example, on \iaArch{}, AFL++'s instruments \texttt{CMP} and \texttt{SUB} instructions, by modifying QEMU's translation stage. This has two main issues: 1) since the instrumentation looks for specific instructions, a separate implementation is required for each architecture, 2) it can result in excessive instrumentation, for example most \texttt{SUB} operations on \iaArch{} are not used for comparisons. CmpLog is not supported in Unicorn.

\vspace{1mm}
\noindent\textbf{CompareCov.}~In \emulatorName{}, integer comparisons are identified using the same algorithm as CmpLog. Once identified, \emulatorName{} injects code that writes to the coverage bitmap for each matching byte before the original comparison operation. For memory comparisons functions, \emulatorName{} searches for the target functions in the program's symbol table and injects instrumentation when a block calling the target function is translated. This allows \emulatorName{'s} instrumentation to be used for statically linked binaries including firmware (as long as the symbol table has not been stripped).

In contrast, AFL++'s implementation for integer comparisons requires identifying architecture specific comparison instructions, like CmpLog instrumentation. Further, for memory comparison functions, it relies on the dynamic linker to replace the original comparison functions with instrumented versions. This approach is unable to support instrumenting statically linked firmware binaries.

\vspace{1.5mm}
\begin{mdframed}[backgroundcolor=black!10,rightline=false,leftline=false,topline=false,bottomline=false,roundcorner=2mm]
\noindent\textbf{Summary.~}\emulatorName{'s} design enables it to satisfy all fuzzing instrumentation requirements in a manner that is simultaneously: i) efficient; ii) avoids new instrumentation requiring direct modification of the emulator internals; and iii) is architecture agnostic. Each instrumentation technique is implemented targeting P-code enabling it to support any ISA. And, as an added benefit, \textit{only knowledge of P-code is adequate} for developing new instrumentation techniques.
\end{mdframed}

\begin{table*}[h!]
\centering
\caption{
Results from different fuzzing instrumentation configurations for the test program. ✓ denotes the bug ID was found at least once within 10 minutes. Each test was repeated 20 times. Shaded grey areas are due to: i)~unsupported fuzzing instrumentation for MIPS and RISC-V in QEMU emulation with AFL++; and ii)~MSP430 ISA being unsupported in QEMU.}
\label{table:bench_portability}

\resizebox{\linewidth}{!}{%
\begin{tabular}{ll|ccccc|ccccc|ccccc|ccccc|ccccc|}
\cline{3-27}
 &  & \multicolumn{5}{c|}{\textsf{\textbf{x86-64}}} & \multicolumn{5}{c|}{\textsf{\textbf{AArch64}}} & \multicolumn{5}{c|}{\textsf{\textbf{MIPS}}} & \multicolumn{5}{c|}{\textsf{\textbf{RISC-V}}} & \multicolumn{5}{c|}{\textsf{\textbf{MSP430}}} \\ \hline
\multicolumn{1}{|l}{Fuzzer} & Instrumentation & \multicolumn{1}{c|}{1} & \multicolumn{1}{c|}{2} & \multicolumn{1}{c|}{3} & \multicolumn{1}{c|}{4} & 5 & \multicolumn{1}{c|}{1} & \multicolumn{1}{c|}{2} & \multicolumn{1}{c|}{3} & \multicolumn{1}{c|}{4} & 5 & \multicolumn{1}{c|}{1} & \multicolumn{1}{c|}{2} & \multicolumn{1}{c|}{3} & \multicolumn{1}{c|}{4} & 5 & \multicolumn{1}{c|}{1} & \multicolumn{1}{c|}{2} & \multicolumn{1}{c|}{3} & \multicolumn{1}{c|}{4} & 5 & \multicolumn{1}{c|}{1} & \multicolumn{1}{c|}{2} & \multicolumn{1}{c|}{3} & \multicolumn{1}{c|}{4} & 5 \\ \hline
\multicolumn{1}{|l}{} & Cov & ✓ & ✓ & - & - & - & ✓ & ✓ & - & - & - & ✓ & ✓ & - & - & - & ✓ & ✓ & - & - & - & ✓ & ✓ & ✓ & - & - \\
\multicolumn{1}{|l}{} & Cov+CmpLog & ✓ & ✓ & ✓ & ✓ & - & ✓ & ✓ & ✓ & ✓ & - & ✓ & ✓ & ✓ & ✓ & - & ✓ & ✓ & ✓ & ✓ & - & ✓ & ✓ & ✓ & ✓ & - \\
\multicolumn{1}{|l}{} & Cov+CompareCov & ✓ & ✓ & ✓ & - & - & ✓ & ✓ & ✓ & - & - & ✓ & ✓ & ✓ & - & - & ✓ & ✓ & ✓ & - & - & ✓ & ✓ & ✓ & - & - \\
\multicolumn{1}{|l}{\multirow{-4}{*}{\emulatorNameBold~(ours)}} & Cov+Context & ✓ & ✓ & - & - & ✓ & ✓ & ✓ & - & - & ✓ & ✓ & ✓ & - & - & ✓ & ✓ & ✓ & - & - & ✓ & ✓ & ✓ & ✓ & - & ✓ \\ \hline
\multicolumn{1}{|l}{} & Cov & ✓ & ✓ & - & - & - & ✓ & ✓ & - & - & - & ✓ & ✓ & - & - & - & ✓ & ✓ & - & - & - & \cellcolor[HTML]{EFEFEF} & \cellcolor[HTML]{EFEFEF} & \cellcolor[HTML]{EFEFEF} & \cellcolor[HTML]{EFEFEF} & \cellcolor[HTML]{EFEFEF} \\
\multicolumn{1}{|l}{} & Cov+CmpLog & ✓ & ✓ & ✓ & ✓ & - & ✓ & ✓ & ✓ & - & - & \cellcolor[HTML]{EFEFEF} & \cellcolor[HTML]{EFEFEF} & \cellcolor[HTML]{EFEFEF} & \cellcolor[HTML]{EFEFEF} & \cellcolor[HTML]{EFEFEF} & \cellcolor[HTML]{EFEFEF} & \cellcolor[HTML]{EFEFEF} & \cellcolor[HTML]{EFEFEF} & \cellcolor[HTML]{EFEFEF} & \cellcolor[HTML]{EFEFEF} & \cellcolor[HTML]{EFEFEF} & \cellcolor[HTML]{EFEFEF} & \cellcolor[HTML]{EFEFEF} & \cellcolor[HTML]{EFEFEF} & \cellcolor[HTML]{EFEFEF} \\
\multicolumn{1}{|l}{\multirow{-3}{*}{\textbf{QEMU}}} & Cov+CompareCov & ✓ & ✓ & ✓ & - & - & ✓ & ✓ & ✓ & - & - & \cellcolor[HTML]{EFEFEF} & \cellcolor[HTML]{EFEFEF} & \cellcolor[HTML]{EFEFEF} & \cellcolor[HTML]{EFEFEF} & \cellcolor[HTML]{EFEFEF} & \cellcolor[HTML]{EFEFEF} & \cellcolor[HTML]{EFEFEF} & \cellcolor[HTML]{EFEFEF} & \cellcolor[HTML]{EFEFEF} & \cellcolor[HTML]{EFEFEF} & \cellcolor[HTML]{EFEFEF} & \cellcolor[HTML]{EFEFEF} & \cellcolor[HTML]{EFEFEF} & \cellcolor[HTML]{EFEFEF} & \cellcolor[HTML]{EFEFEF} \\ \hline
\end{tabular}}
\end{table*}

\vspace{-0.1cm}
\subsection{Fuzzing Frontend Integration}\label{integration-with-afl}

Modern grey-box fuzzing frameworks consists of two main components: the \textit{frontend} which handles input generation, input scheduling, hang detection and crash deduplication, and the \textit{backend} which manages program execution, crash monitoring, and instrumentation. Emulation-based fuzzers utilize emulators as the fuzzing backend allowing for \textit{binary-only} and \textit{cross-architecture} fuzzing.
\emulatorName{} is a new fuzzing backend, therefore, we make our emulator compatible with an existing fuzzing framework: AFL++~\cite{aflplusplus2020} to avoid implementing a new frontend. AFL++ is a state-of-the-art fuzzing framework derived from the well-known American Fuzzy Lop (AFL)~\cite{afl2010} project, with general improvements, and support for additional fuzzing techniques. \emulatorName{} integrates with AFL++ using the forksever interface also used by AFL++'s QEMU-mode.

\section{Evaluation}\label{evaluation}

\noindent\textbf{Settings.~}Unless otherwise specified, all experiments were carried out with \aflVersion{} as the fuzzing frontend on an \benchmarkCPU{}
restricted to a single core. All AFL++ settings were kept as default, except to enable instrumentation as needed and to adjust the timeout for hang detection.

\vspace{1mm}
\noindent\textbf{Experiments.~}We design our experimental regime to answer five specific questions articulated in Section~\ref{portability-of-instrumentation}-\ref{sec:performance-testing}.

\subsection{Is \emulatorName{'s} Instrumentation Portable Across Diverse ISAs?}\label{portability-of-instrumentation}

To ensure that architecture-agnostic instrumentation implemented in our emulator is operational across a range of architectures, we designed a test program, shown in Listing~\ref{listing:test-instrumentation}, that consists of 5 synthetic bugs designed to test specific instrumentation.

\begin{listing}[tbhp!]
\begin{minted}[framesep=2mm,fontsize=\footnotesize]{c}
void test_instrumentation(char* buf) {
    // (1) comparison against a single byte in the input
    if (buf[0] == '%') {
        crash(1);
    }
    // (2) Multiple comparison against single bytes of the input.
    if (buf[0] == 'i' && buf[1] == 'x' &&
        buf[2] == 'S' && buf[3] == 'D') {
        crash(2);
    }
    // (3) A single comparison against multiple input bytes.
    if (*(u32*)buf == *(u32*)"wzfc") {
        crash(3);
    }
    // (4) A multi-byte comparison across a function call.
    if (compare(buf, "dGlIHF1W") == 0) {
        crash(4);
    }
    // (5) Saturate coverage then compare.
    saturate_compare2_cov();
    u32 tmp = *((u32*)buf) ^ 0x46092d5f;
    if (compare2(tmp, 0x7451496b)) {
        crash(5);
    }
}
\end{minted}
\caption{Test program used for evaluating instrumentation.} 
\label{listing:test-instrumentation}
\end{listing}

We evaluate the portability of \emulatorName{'s} instrumentation by fuzzing the test program compiled for 5 different architectures. For architectures with Linux support, we configure the program to read the input from \texttt{stdin}. For MSP430, the program reads from a peripheral mapped to the fuzzing input. After compiling the binary for each architecture, we manually verified that the machine code of output binary behaves as expected. As a baseline we compare against AFL++'s QEMU-mode when instrumentation is supported for the guest architecture. For each fuzzing configuration, we perform 20 trials for a maximum of 10 minutes starting with an uninformed seed. The results from this experiment are shown in Table~\ref{table:bench_portability}.

Both \textit{Bug1} and \textit{Bug2} are discoverable with code coverage alone, so are found by all fuzzing configurations. \textit{Bug3} requires additional instrumentation to be found so can only be found when one of the two comparison instrumentation techniques is enabled, except for the MSP430 binary. With CmpLog, the fuzzer can find a solution via an input-to-state mutation directly replacing the incorrect value with a correct one, with CompareCov enabled, the comparison is split into byte-level comparisons, and the fuzzer observes incremental coverage feedback similar to \textit{Bug2}. Since MSP430 is a 16-bit architecture, the compiler splits the 32-bit comparison into two 16-bit comparisons allowing the fuzzer to eventually find the crashing input for \textit{Bug3} without additional instrumentation. \textit{Bug4} evaluates the fuzzers ability to solve memory comparison functions so is only discovered when CmpLog instrumentation enabled, which generally finds the crashing input within seconds. CompareCov fails to find the bug, since \texttt{compare} is not a standard comparison function and is therefore not instrumented by CompareCov. CmpLog is only partially implemented for QEMU on AArch64 (function calls are not instrumented) so fails to find \textit{Bug4}. \textit{Bug5} tests the fuzzer's ability to find a bug in a function where code coverage is saturated by a previous call, so is only found when context sensitive branch coverage is enabled, which only \emulatorName{} supports on all architectures.

\vspace{1.5mm}
\begin{mdframed}[backgroundcolor=black!10,rightline=false,leftline=false,topline=false,bottomline=false,roundcorner=2mm]
    \textbf{Summary} The test program binaries for five different ISAs provide empirical evidence that the architecture agnostic instrumentation implementation of the different instrumentation techniques in \emulatorName{} is both effective and portable across architectures.
\end{mdframed}

\begin{figure*}[t!]
\centering
\includegraphics[width=\linewidth]{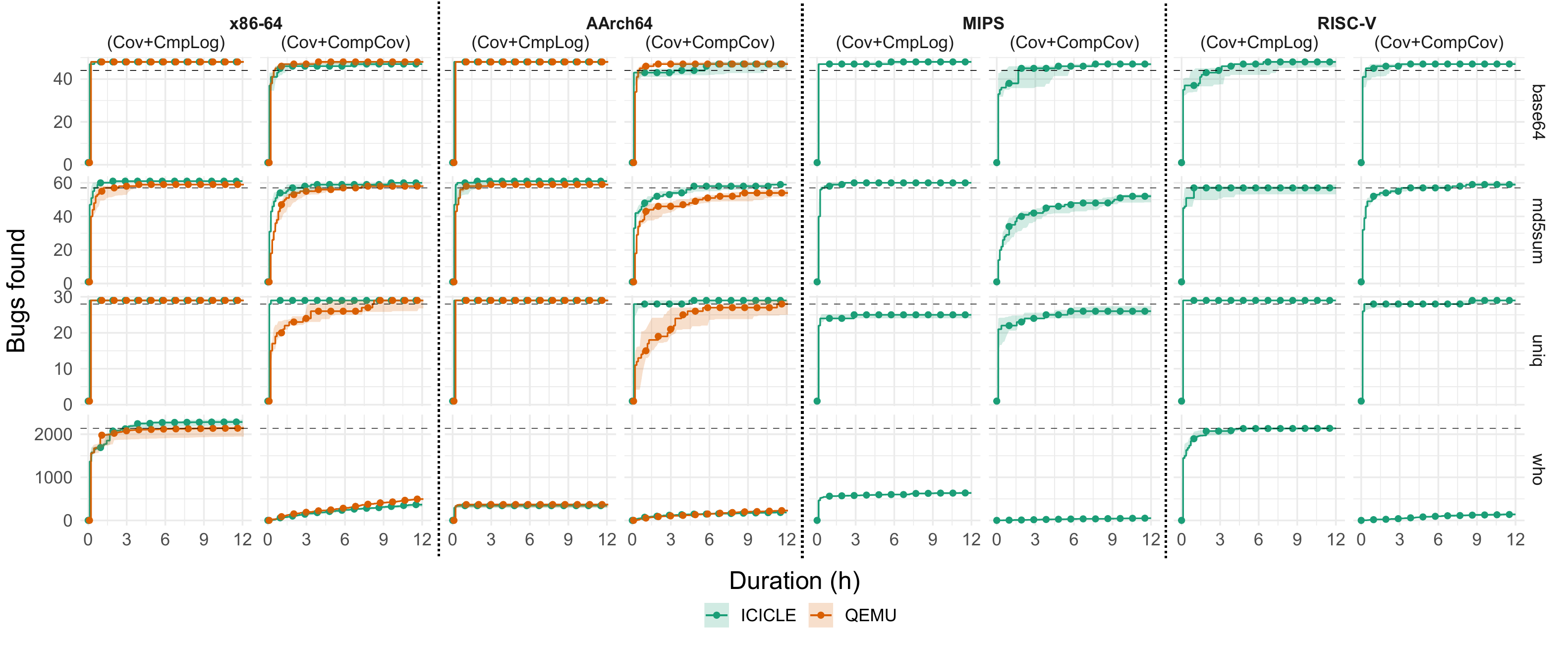}
\caption{LAVA-M bugs found over time in each binary. The solid line represents the median number of bugs found, the shaded area represents the min/max coverage across all trials, and the black dotted lines represent the number of bugs listed in the LAVA-M paper (\textit{Note}: it is well known that it is possible to trigger additional bugs other than specified in the original paper).}
\label{fig:lava_bugs}
\end{figure*}

\subsection{Is Architecture-Agnostic Instrumentation as Effective as Existing Architecture-Specific Implementations?}\label{sec:lava-m-benchmarks}

LAVA-M~\cite{lava2016} is a widely used set of binaries for evaluating and benchmarking fuzzers. It consists of four binaries from GNU coreutils~\cite{gnucoreutils} each injected with synthetic bugs. While the injected bugs are not representative of typical real-world vulnerabilities~\cite{fuzzeval2018}, previous work has demonstrated that these bugs are difficult to find with code coverage only but can be found with by instrumenting comparison operations~\cite{redqueen2019, greyone2020, angora2018}. This naturally lends itself to assessing \emulatorName{'s} architecture-agnostic implementation of CmpLog and CompareCov instrumentation.

Using AFL++ as the frontend, we evaluate the bug discovery capability of \emulatorName{} across four different ISAs (\iaArch{}, AArch64, MIPS, and RISC-V), the first two of which we compare against QEMU\footnote{We compare with QEMU not Unicorn, since Unicorn cannot directly execute Linux binaries. Additionally, since neither CmpLog nor CompareCov instrumentation are supported in AFL++'s QEMU-mode for RISC-V and MIPS, we only evaluate these architectures with \emulatorName{}}. We also evaluate both QEMU and \emulatorName{} on \iaArch{} with code coverage only as a baseline. For each of the injected bugs, a unique ID is written to \texttt{stdout} whenever the bug is triggered. Therefore, we can verify each crash by running the \iaArch{} version of the binary on the host machine then checking for unique bug IDs in \texttt{stdout}. We perform 5 trials for each fuzzing configuration running for 12 hours each, starting with the same two initial seeds as Section~\ref{portability-of-instrumentation}. Figure~\ref{fig:lava_bugs} shows the bugs found over-time with instrumentation enabled and Table~\ref{table:lava_m_results} in the Appendix details the total bugs found for every architecture, instrumentation, and emulator configuration.

With code coverage alone almost no bugs are found by either emulator in any of the binaries. Both comparison instrumentation techniques allow most bugs to be found, with CmpLog finding bugs significantly faster than CompareCov in several cases. \emulatorName{'s} results closely match QEMU results for both AArch64 and \iaArch{}, which supports our claim that \emulatorName{'s} instrumentation is as effective as the architecture-specific approach employed by AFL++'s QEMU-mode.
On the two additional architectures tested with \emulatorName{} both instrumentation techniques continue to be effective. However, the results for the MIPS version of \texttt{uniq} are slightly worse, this is caused by differences in the memory layout (MIPS uses a 32-bit address space, while the other architectures are 64-bit), which causes issues when replaying the crashing input on the \iaArch{} host.

The differences in the number of crashes found for \texttt{who} binary across architecture is caused caused by platform specific behaviour in the program itself. The fuzz input is parsed as a \texttt{utmpx} structure, however the layout of the fields within the structure is different across architectures. This can cause certain bugs to become unreachable, and  causes issues when we attempt to replay the crashing inputs on the \iaArch{} version of the binary in order to verify the crash IDs. Additionally, the binary frequently crashes before a bug ID is flushed to \textit{stdout} (caused by internal buffering), which prevents us from obtaining the bug ID from the original execution. Notably, all bugs reported and discovered are those reproduced on both the guest architecture and the host (\iaArch{}). This is additional evidence of the importance of \textit{binary-only} and \textit{cross-architecture} fuzzing; even when source code is available, program behaviour can differ on between architectures.

\vspace{1.5mm}
\begin{mdframed}[backgroundcolor=black!10,rightline=false,leftline=false,topline=false,bottomline=false,roundcorner=2mm]
    \textbf{Summary} Discovering LAVA-M benchmark bugs require a specific operational capability from instrumentation techniques to solve comparison operations; namely CompCov or CompLog. \emulatorName{'s} results closely match QEMU results for both AArch64 and \iaArch{}, supporting our claim that \emulatorName{'s} instrumentation is as effective as the architecture-specific approach employed by AFL++'s QEMU-mode. On MIPS and RISC-V architectures (where AFL++'s QEMU-mode does not support the necessary instrumentation) both instrumentation techniques tested with \emulatorName{} continue to be effective.
\end{mdframed}

\subsection{Can \emulatorName{} Be Used to Implement and Enhance State-of-the-Art Firmware Fuzzing Techniques?}\label{sec:arm-fuzzing}

Fuzzware~\cite{fuzzware2022} is a recent state-of-the-art fuzzing framework for analyzing ARM firmware binaries. Fuzzware extends Unicorn to instrument and execute ARM firmware. We replace Unicorn with \emulatorName{} to evaluate \emulatorName{'s} ability to support state-of-the-art firmware fuzzing. We then tested our modified version ({\sc Fuzzware}-\emulatorName{}) by attempting to reproduce {\sc Fuzzware}'s results on the 10 \textit{real-world} binaries used in the P$^2$IM~\cite{p2im2020} firmware set as they were evaluated extensively by~\cite{p2im2020,uemu2021} and {\sc Fuzzware}. Importantly, since \emulatorName{'s} instrumentation is portable we can support \textit{additional instrumentation} when fuzzing ARM firmware. In particular, we perform additional tests with CompareCov instrumentation enabled to allow for better comparison solving\footnote{We did not test with CmpLog, since effective use of the instrumentation requires additional integration with fuzzing frontend, unsupported by {\sc Fuzzware}.}. We followed the same experimental setup for {\sc Fuzzware} as described in the original paper (we used the same number of trials, seeds and run time duration).

We were able to successfully rediscover all 16 of the bugs found by {\sc Fuzzware}, and, with CompareCov enabled, {\sc Fuzzware}-\emulatorName{} was able to find an additional bug in the Console binary \textit{not reported} by any prior work\footnote{Crashing inputs for each of the discovered bugs are available in our GitHub repository.}. As part of the \texttt{rtc settime} command, the firmware reads a date from the user in the form \texttt{YYYY-MM-DD HH:MM:SS} without checking whether the parsed date is valid. This causes an out-of-bounds access when the name of the month is resolved using a lookup table. Since reaching this bug requires first solving a string comparison to reach the \texttt{rtc} handler, then solving a second string comparison for the \texttt{settime} subcommand, we believe the added instrumentation was critical to finding this bug.

\emulatorName{} also found an additional crash in the Soldering Iron binary. At high temperatures, rendering the heat indicator causes the buffer allocated for the LCD screen to overflow. However, after further analysis we discovered the maximum temperature is \textit{restricted} in software, indicating that the bug is a false-positive caused by {\sc Fuzzware}'s peripheral modelling strategy.

\begin{figure*}[ht]
\centering
\includegraphics[width=\linewidth]{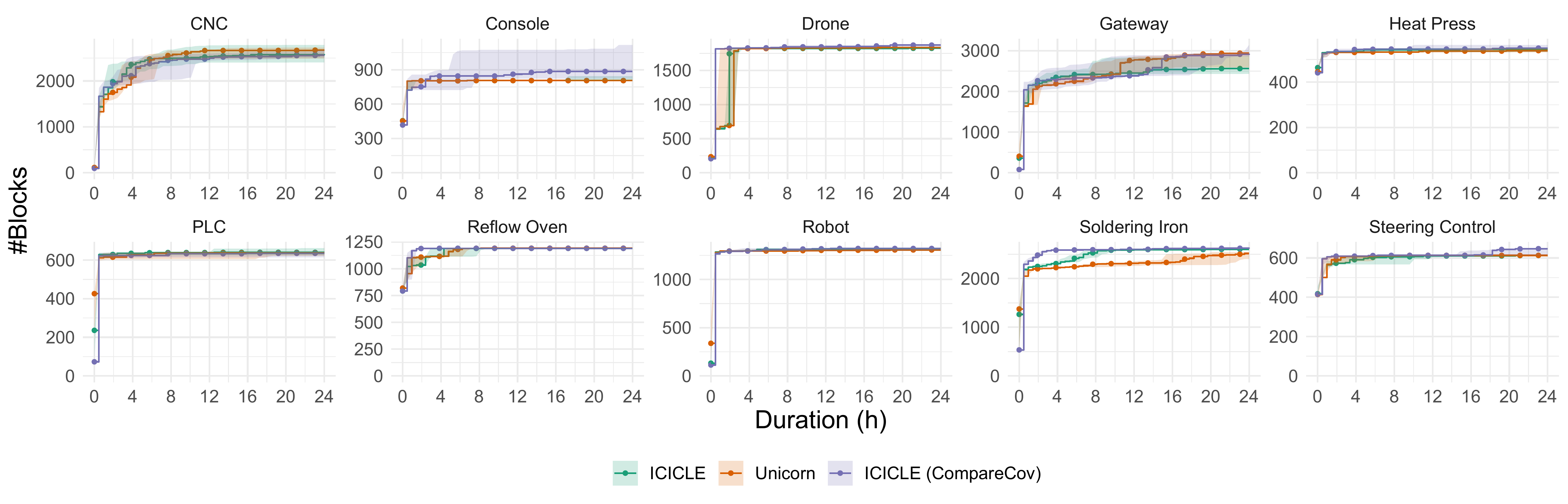}
\caption{Block coverage over time for the \textit{real-world} ARM firmware using the two different emulators and CompareCov instrumentation supported in \emulatorName{}. The solid line represents the median coverage of 5 runs, and the shaded area represents min/max coverage.}
\label{fig:fuzzware-blocks}
\end{figure*}

In addition to reproducing the bugs, we also investigated whether {\sc Fuzzware}-\emulatorName{} is able to maintain the same level of block coverage as the original implementation. The results are shown in Figure~\ref{fig:fuzzware-blocks}. For almost all the evaluated binaries we achieve almost identical block coverage to {\sc Fuzzware}, with some small differences of which we manually investigated.
With CompareCov enabled, {\sc Fuzzware}-\emulatorName{} achieves higher coverage in two of the binaries: Console, and Steering Control. The higher coverage in Console corresponds to reaching different command handlers that are dispatched based on string comparisons, including the sub-commands of the \texttt{rtc} handler that contains the bug discussed above. Similarly, Steering Control contains two commands, that are triggered when the matching string is read by the firmware (\texttt{"steer"}, and \texttt{"motor"}). CompareCov enables \emulatorName{} to generate inputs containing the command strings, and thus is able to reach additional code. The discrepancies in the Gateway and Soldering Iron binaries are caused by high variance between fuzzing runs, running additional trials would likely remove any discrepancies.

\vspace{1.5mm}
\begin{mdframed}[backgroundcolor=black!10,rightline=false,leftline=false,topline=false,bottomline=false,roundcorner=2mm]
    \textbf{Summary} \emulatorName{} is a robust emulator capable of supporting the current state-of-the-art ARM firmware fuzzer, {\sc Fuzzware}. We discovered all 16 known bugs. \emulatorName{} provides a direct substitute for Unicorn with the added advantage of additional, architecture agnostic instrumentation shown to be effective at \textit{improving coverage} and \textit{discovering 2 new bugs} in real-world firmware not reported by fuzzing efforts in prior work.
\end{mdframed}

\subsection{Can \emulatorName{} Discover Bugs in Binaries in an ISA Currently Not Supported by Emulation-Based Fuzzers?}\label{sec:msp430-fuzzing}

To demonstrate the architecture-independent benefits of our prototype emulator, we investigate fuzzing firmware written for MSP430 microcontrollers. FiE~\cite{fie2013}, is the only prior study that attempted to find bugs in MSP430 firmware. However, FiE requires C source-code and therefore does not support manually written assembly code (which is common in larger firmware) and is incapable \textit{binary-only} fuzzing.
Further, MSP430 firmware is not supported by any existing emulation-based fuzzing framework\footnote{\cite{msp430qemu2016} is a fork of QEMU adding MSP430 support, however it is outdated, not integrated with any fuzzing framework, and does not support MSP430 CPUX extensions.}, and therefore presents a compelling use case for fuzzing with \emulatorName{}.

\begin{figure}[!htbp]
\centering
\includegraphics[width=0.7\linewidth]{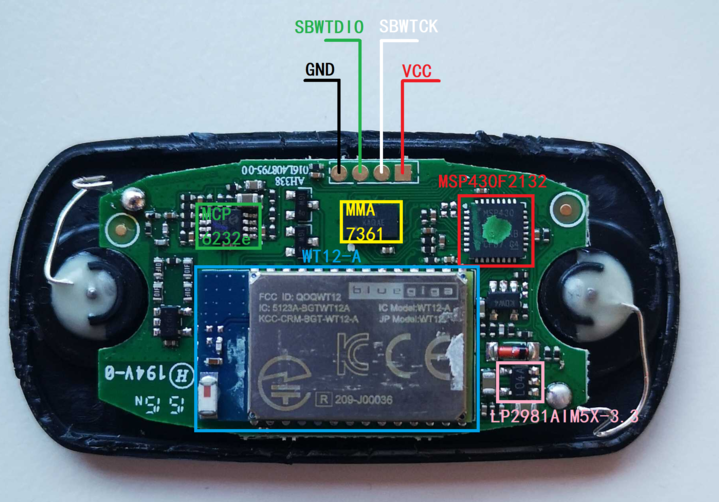}
\caption{The internals of a dismantled Polar Heart Rate Tracker.}
\label{fig:polar-photo}
\end{figure}

\begin{table}[]
\caption{Discovered bugs in MSP430 \textit{real-world} binaries.}\label{table:msp430-bugs}
\resizebox{\linewidth}{!}{%
\begin{tabular}{l|l}
\textbf{Firmware}  & \textbf{Bug description (PoCs \& stack traces on GitHub)}                          \\ \hline
Goodwatch          & Incorrect comparison when writing to log buffer.  \\
Goodwatch          & Buffer overflow when handling zero length packet. \\
Goodwatch          & Stack overflow in RNG generation.                 \\
Goodwatch          & Out-of-bounds access in OOK keypress.             \\
Goodwatch          & Out-of-bounds access in Stopwatch.                \\
H4\_PacketProtocol & Unchecked Interface Index in Get Descriptor.      \\
H4\_PacketProtocol & Buffer overflow in Set Report.
\end{tabular}}
\end{table}

Similar to existing monolithic firmware fuzzing approaches~\cite{p2im2020, fuzzware2022}, we handle peripheral accesses for MSP430 firmware by reading them from the fuzzer input. We found this highly effective at finding bugs. We selected 3 different firmware to evaluate. First, inspired by FiE, we evaluated the USB SDK provided as part of TI's MSP430 USB Developers Package\footnote{\url{https://www.ti.com/tool/MSP430USBDEVPACK} version 5.20.06.03} using example programs provided as part of the development package (H4\_PacketProtocol) as a harness. Second, we compiled an unmodified version of the Goodwatch~\cite{goodwatch2021} firmware\footnote{commit: c8859f845fccf56585a127059b1d1b825b381673}, a hardware and firmware replacement for Casio calculator watches based on the CC430 MCU (a MSP430 CPU with an integrated RF transceiver) as that was the highest-ranking application on GitHub's trending page for MSP430. Additionally, we investigated \emulatorName{'s} ability to test closed-source firmware by extracting the firmware off a commercial medical device, a Polar heart rate tracker, containing a MSP430F2132~MCU as shown in Figure~\ref{fig:polar-photo}.

The block coverage results are summarized Table~\ref{table:msp430-blocks}. After triaging the results, we identified two unique bugs H4\_PacketProtocol and 5 unique bugs in the Goodwatch firmware and 3 additional crashes related to debugging features. While no bugs were discovered for the Polar heart rate tracker, the fuzzer reached almost all blocks in the firmware. The bugs discovered by \emulatorName{} are summarized in Table~\ref{table:msp430-bugs}, and for each bug discovered we provide input files and a detailed crash analysis in our GitHub repository\footnote{\url{https://github.com/icicle-emu/icicle/tree/main/crash-analysis/msp430}}.

\vspace{1.5mm}
\begin{mdframed}[backgroundcolor=black!10,rightline=false,leftline=false,topline=false,bottomline=false,roundcorner=2mm]
    \textbf{Summary} MSP430 firmware fuzzing is not supported by existing emulation-based fuzzing frameworks. Case studies with \emulatorName{} and its suite of architecture agnostic instrumentation discovered seven undiscovered software bugs in two (USB SDK--H4\textunderscore PacketProtocol, and Goodwatch) of the three tested MSP430 binaries.
\end{mdframed}

\begin{table}[]
\caption{Block coverage (\#BB) for the real-world MSP430 binaries with and without CmpLog instrumentation enabled. Avg represents the median coverage achieved after 24 hours in 5 trials.}\label{table:msp430-blocks}
\resizebox{\linewidth}{!}{%
\begin{tabular}{|l|c|l|rrr|}
\hline
\textbf{Firmware} & \textbf{\begin{tabular}[c]{@{}c@{}}\#BB \\ total\end{tabular}} & \textbf{Instrumentation} & \multicolumn{1}{c}{\textbf{\begin{tabular}[c]{@{}c@{}}\#BB \\ min\end{tabular}}} & \multicolumn{1}{c}{\textbf{\begin{tabular}[c]{@{}c@{}}\#BB \\ avg\end{tabular}}} & \multicolumn{1}{c|}{\textbf{\begin{tabular}[c]{@{}c@{}}\#BB \\ max\end{tabular}}} \\ \hline
\multirow{2}{*}{Goodwatch} & \multirow{2}{*}{3263} & Cov & \multicolumn{1}{l}{2336} & 2362 & 2441\\
 &  & Cov+CmpLog & \textbf{2438} & \textbf{2503} & \textbf{2526} \\ \hline
\multirow{2}{*}{\begin{tabular}[c]{@{}l@{}}H4 Packet\\ Protocol\end{tabular}} & \multirow{2}{*}{925} & Cov & \textbf{819} & 821 & 891\\
 &  & Cov+CmpLog & 813 & \textbf{910} & \textbf{914} \\ \hline
\multirow{2}{*}{\begin{tabular}[c]{@{}l@{}}Heart Rate \\ Tracker\end{tabular}} & \multirow{2}{*}{744} & Cov & 679 & 679 & 716 \\
 &  & Cov+CmpLog & \textbf{680} & \textbf{717} & \textbf{718} \\ \hline
\end{tabular}}
\end{table}

\subsection{How Does \emulatorName{} Perform in Fuzz Testing?}\label{sec:performance-testing}

\begin{figure}[!htbp]
\centering
\includegraphics[width=\linewidth]{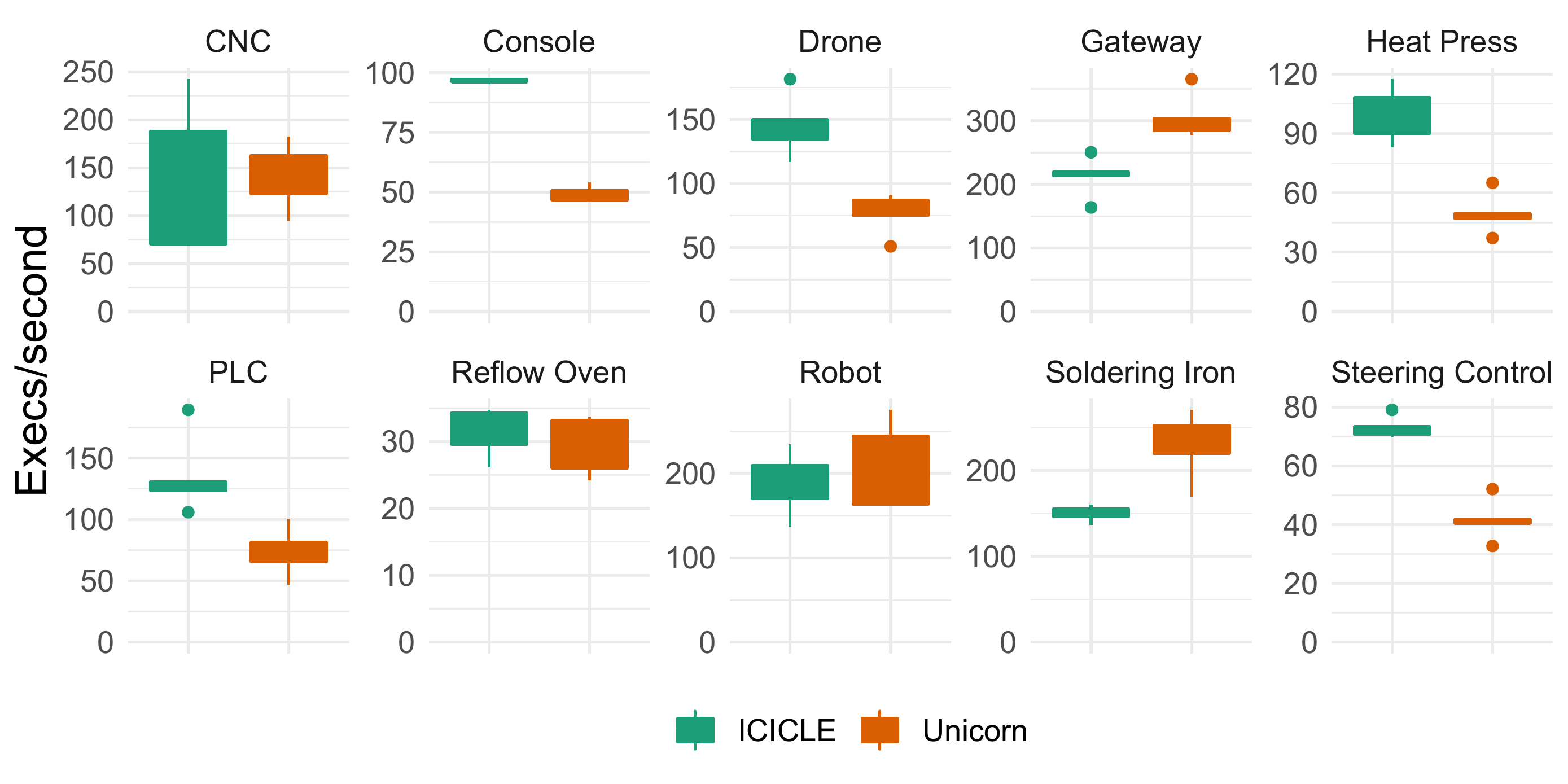}
\caption{\emulatorName{} and Unicorn performance comparison when integrated with the state-of-the-art fuzzer, {\sc Fuzzware}.}
\label{fig:fuzzware-perf}
\end{figure}

In the development of \emulatorName{},  we made efforts to ensure that \emulatorName{} has good performance in general. We compared \textit{fuzz test execution speed} of \emulatorName{} with Unicorn (emulator) employed by the sate-of-the-art fuzzer, {\sc Fuzzware}, on the P$^2$IM dataset~\cite{p2im2020} and summarise the results in Figure~\ref{fig:fuzzware-perf}.

\vspace{1.5mm}
\begin{mdframed}[backgroundcolor=black!10,rightline=false,leftline=false,topline=false,bottomline=false,roundcorner=2mm]
    \textbf{Summary} \emulatorName{} has approximately the same performance as Unicorn for fuzzing monolithic firmware binaries.
\end{mdframed}

\section{Discussion and Limitations}\label{discussion-and-limitations}

Although we have taken the first steps to re-think and re-design an emulation framework to directly support fuzzing requirements, and instrumentation development and testing, the current implementation is not without limitations. The released emulator prototype was primarily designed for CPU ISA emulation, similar to the goals of the Unicorn project. As a result, Linux emulation is minimal, and more complex hardware emulation required for full-system emulation (e.g., page-table emulation) is not currently supported.

\subsection{Emulator Correctness}\label{emulator-testing}

In emulation-based fuzzing, since the program not executed on the original hardware, there is a risk that any crashes discovered could be caused by emulation issues, not bugs in the target program. To reduce the chance of emulation bugs in \emulatorName{}, first, we employ a \textit{differential testing} strategy, similar to other widely used approaches for testing CPU emulators~\cite{emufuzzer2009, kemufuzz2010, pokeemu2012, virtcpuvalidation2015, fastpokeemu2018, verifyingiss2019}. Second, we manually investigated any crashes discovered in benchmark evaluations and ensure they are caused by program bugs.

\subsection{Performance}\label{sec:performance}

In the development of \emulatorName{}, while we made efforts to ensure that \emulatorName{} has good performance in general, there are several additional optimizations possible. The current implementation of \emulatorName{} has demonstrably similar performance to Unicorn (see Figure~\ref{fig:fuzzware-perf}). Although a  direct performance comparison against QEMU is desirable, it is more difficult.  \emulatorName{} implements a forkserver similar to AFL++'s persistent mode, however we run AFL++'s QEMU-mode without this feature since (currently) persistent mode requires a significant amount of manual effort to set up for each binary (notably, \emulatorName{}'s implementation is automatic for Linux binaries). This results in \emulatorName{} performing significantly faster for small binaries. Other the other hand, since \emulatorName{} always translates memory accesses in software (like Unicorn) while AFL++'s QEMU-mode can utilize hardware address translation when running Linux user-space binaries on a Linux host, we expect a speedup for QEMU for larger Linux binaries.

\section{Related work}\label{related-work}

\noindent\textbf{Improving emulation-based fuzzing.~}
There has been some effort in improving QEMU and Unicorn for fuzzing, including, improving runtime performance~\cite{improvingaflqemu2018, aflplusplus2020}, enabling support for full-system emulation of Linux-based firmware~\cite{triforceafl2016, firmadyne2016, firmafl2019, equafl2022, qlinggithub2022}, and extending the emulator to support additional analysis such as taint tracking and symbolic execution~\cite{bitblaze2008, symqemu2021}.

\vspace{2mm}
\noindent\textbf{Binary-only fuzzing.~} Without access to source-code it is challenging to use fuzzing techniques that rely on instrumentation, since the simplest approach using compiler-based code injection, is not possible. Fuzzers that support targets without source-code are known \textit{binary-only} fuzzers. Emulation-based approaches are one solution, however there are several other alternatives.

Virtualisation/hardware-assisted approaches (e.g., kAFL~\cite{kafl2017}, and NyX~\cite{nyx2020,nyxnet2022}) use a variety of hardware features to implement fuzzing instrumentation. Since they require additional hardware support some instrumentation cannot be easily implemented, and firmware fuzzing is not supported. Static rewriting techniques (e.g., Retrowrite~\cite{retrowrite2020}, Datalog Disassembly~\cite{datalogdisasm2020}, {\sc Zafl}~\cite{zafl2021}) disassemble a binary, inject instrumentation, then reassemble the binary. This can enable close to compiler-level instrumentation performance, however the complexity involved in the rewriting process often results in correctness issues, typically firmware binaries are not well supported, and static rewriting cannot be used cross-architecture fuzzing. While, dynamic instrumentation tools (e.g., DynamoRIO~\cite{dynamorio2012}, PIN~\cite{pin2005}, CMU BAP~\cite{bap2011}, Valgrind~\cite{valgrind2007}), share significant similarities to emulation-based approaches, they are more restrictive than full emulators and are unable to support firmware fuzzing.

\vspace{2mm}
\noindent\textbf{Embedded system fuzzing.~} Fuzzing embedded systems and IoT devices is difficult because we cannot avoid dealing with hardware and peripheral interactions since it might represent a majority of the code we are trying to test. As a result, emulation-based fuzzers need to support more than just CPU emulation. Past work has extended either QEMU or Unicorn to support firmware fuzzing through, hardware-in-the-loop approaches~\cite{inception2018}, peripheral modeling~\cite{pretender2019, fie2013, p2im2020, uemu2021}, or emulating the hardware abstraction layer~\cite{halucinator2020}. More recently, \textsc{Fuzzware}~\cite{fuzzware2022} investigated techniques to automatically generate peripherals models using local symbolic execution. More recently, \textsc{Ember-IO}~\cite{emberio2023} removed the need for peripherals models and exploited commonalities in the way hardware behaves to make better use of input data while minimizing the impact of poor coverage feedback resulting from excess coverage caused by interrupts. Both \textsc{Fuzzware} and \textsc{Ember-IO} demonstrated results outperforming prior firmware fuzzing approaches. Notably, aside from FiE (which is source-based and only targets MSP430), all these approaches only evaluate firmware written for the ARM architecture and either employ Unicorn or QEMU. \emulatorName{} makes it easier to fuzz multiple architectures, which we hope will assist in increasing the architecture diversity in future work. MetaEmu~\cite{metaemu2022} also uses SLEIGH to support emulating multiple architectures, however, MetaEmu's primary goal is to support general dynamic binary analysis, not high-performance fuzzing, and as such lacks a JIT-based execution model, instead focusing on supporting flexible execution modes.

\section{Conclusion}
Emulation-based fuzzing techniques, supported by effective instrumentation, are highly flexible and are the only method for  \textit{cross-architecture} fuzzing. For historical reasons, emulators used in existing emulation-based fuzzing frameworks were not designed for fuzzing and has made it difficult to meet fuzzing specific needs such as implementing advanced instrumentation techniques supporting a  design-build-and-test once-only paradigm across multiple ISAs and implementing fuzzing specific optimizations.

We designed and implemented a new multi-architecture emulation framework for fuzzing. Within our framework, we implemented four different architecture agnostic instrumentation techniques and demonstrated that a single architecture-independent implementation is effective across multiple architectures. Our emulation platform is extremely flexible, supporting a wide range of ISAs, especially significant in fuzzing firmware in embedded systems and IoT devices. This was demonstrated by discovering 7 new bugs in ARM firmware by integrating with the state-of-the-art ARM firmware fuzzer, {\sc Fuzzware} and fuzzing firmware for MSP430 ISAs---an unsupported target by existing emulation-based fuzzers.

\section{Data Availability Statement}
We uploaded artifacts to \url{https://github.com/icicle-emu/icicle}, including source code, a README guide for users, proof-of-crash (PoC) inputs and stack traces for all discovered bugs.

\section*{Acknowledgements}

The work has been supported by the Australian Government's Research Training Program Scholarship (RTPS) and Cyber Security Research Centre Limited whose activities are partially funded by the Australian Government’s Cooperative Research Centres Programme.

\section*{Appendix}

\begin{table}[!htbp]
\centering
\caption{Total LAVA-M bugs found within 5 trials of each configuration over 12 hours. The results demonstrate the effectiveness of the architecture agnostic instrumentation as well as the benefits from making available state-of-the-art instrumentation across architectures with ease.}
\label{table:lava_m_results}
\resizebox{\linewidth}{!}{%
\begin{tabular}{|c|ll|rrrr|}
\hline
\multirow{2}{*}{ISA} & \multirow{2}{*}{Fuzzer} & \multirow{2}{*}{Instrumentation} & \multicolumn{4}{c|}{LAVA-M bugs found} \\ \cline{4-7}
                         &                         &                & base64 & md5sum & uniq & who  \\ \hline
\multirow{6}{*}{x86-64}  & \multirow{3}{*}{ICICLE} & Cov            & 4      & 0      & 1    & 0    \\
                         &                         & Cov+CmpLog     & 48     & 61     & 29   & 2419 \\
                         &                         & Cov+CompareCov & 48     & 60     & 29   & 1002 \\ \cline{2-7}
                         & \multirow{3}{*}{QEMU}   & Cov            & 0      & 0      & 1    & 0    \\
                         &                         & Cov+CmpLog     & 48     & 59     & 29   & 2356 \\
                         &                         & Cov+CompareCov & 48     & 59     & 29   & 1256 \\ \hline
\multirow{6}{*}{AArch64} & \multirow{3}{*}{ICICLE} & Cov            & 0      & 0      & 2    & 0 \\
                         &                         & Cov+CmpLog     & 48     & 61     & 29   & 375  \\
                         &                         & Cov+CompareCov & 48     & 60     & 29   & 357  \\ \cline{2-7}
                         & \multirow{3}{*}{QEMU}   & Cov            & 2      & 0      & 0    & 0  \\
                         &                         & Cov+CmpLog     & 48     & 59     & 29   & 389  \\
                         &                         & Cov+CompareCov & 48     & 59     & 29   & 381  \\ \hline
\multirow{4}{*}{MIPS}    & \multirow{3}{*}{ICICLE} & Cov            & 1      & 0      & 1    & 0  \\
                         &                         & Cov+CmpLog     & 48     & 61     & 28   & 799  \\
                         &                         & Cov+CompareCov & 48     & 60     & 29   & 148   \\ \cline{2-7}
                         & \multirow{1}{*}{QEMU}   & Cov            & 0      & 0      & 0    & 0  \\ \hline
\multirow{4}{*}{RISC-V}  & \multirow{3}{*}{ICICLE} & Cov            & 1      & 0      & 0    & 0 \\
                         &                         & Cov+CmpLog     & 48     & 59     & 29   & 2327 \\
                         &                         & Cov+CompareCov & 48     & 60     & 29   & 472  \\ \cline{2-7}
                         & \multirow{1}{*}{QEMU}   & Cov            & 1      & 0      & 0    & 0  \\ \hline
\end{tabular}}
\end{table}


    \bibliographystyle{ACM-Reference-Format}
    \bibliography{references.bib}


\begin{thebibliography}{69}


\ifx \showCODEN    \undefined \def \showCODEN     #1{\unskip}     \fi
\ifx \showDOI      \undefined \def \showDOI       #1{#1}\fi
\ifx \showISBNx    \undefined \def \showISBNx     #1{\unskip}     \fi
\ifx \showISBNxiii \undefined \def \showISBNxiii  #1{\unskip}     \fi
\ifx \showISSN     \undefined \def \showISSN      #1{\unskip}     \fi
\ifx \showLCCN     \undefined \def \showLCCN      #1{\unskip}     \fi
\ifx \shownote     \undefined \def \shownote      #1{#1}          \fi
\ifx \showarticletitle \undefined \def \showarticletitle #1{#1}   \fi
\ifx \showURL      \undefined \def \showURL       {\relax}        \fi
\providecommand\bibfield[2]{#2}
\providecommand\bibinfo[2]{#2}
\providecommand\natexlab[1]{#1}
\providecommand\showeprint[2][]{arXiv:#2}

\bibitem[\protect\citeauthoryear{Alliance}{Alliance}{2021}]%
        {craneliftgithub2021}
\bibfield{author}{\bibinfo{person}{Bytecode Alliance}.}
  \bibinfo{year}{2021}\natexlab{}.
\newblock \bibinfo{title}{Cranelift Code Generator}.
\newblock
  \bibinfo{howpublished}{\url{https://github.com/bytecodealliance/wasmtime/tree/main/cranelift}}.
\newblock


\bibitem[\protect\citeauthoryear{Amit, Tsafrir, Schuster, Ayoub, and
  Shlomo}{Amit et~al\mbox{.}}{2015}]%
        {virtcpuvalidation2015}
\bibfield{author}{\bibinfo{person}{Nadav Amit}, \bibinfo{person}{Dan Tsafrir},
  \bibinfo{person}{Assaf Schuster}, \bibinfo{person}{Ahmad Ayoub}, {and}
  \bibinfo{person}{Eran Shlomo}.} \bibinfo{year}{2015}\natexlab{}.
\newblock \showarticletitle{Virtual CPU Validation}. In
  \bibinfo{booktitle}{\emph{Proceedings of the 25th Symposium on Operating
  Systems Principles}} \emph{(\bibinfo{series}{SOSP '15})}.
  \bibinfo{pages}{311--327}.
\newblock
\urldef\tempurl%
\url{https://doi.org/10.1145/2815400.2815420}
\showDOI{\tempurl}


\bibitem[\protect\citeauthoryear{Andronidis and Cadar}{Andronidis and
  Cadar}{2022}]%
        {snapfuzz2022}
\bibfield{author}{\bibinfo{person}{Anastasios Andronidis} {and}
  \bibinfo{person}{Cristian Cadar}.} \bibinfo{year}{2022}\natexlab{}.
\newblock \showarticletitle{SnapFuzz: High-Throughput Fuzzing of Network
  Applications}. In \bibinfo{booktitle}{\emph{Proceedings of the 31st ACM
  SIGSOFT International Symposium on Software Testing and Analysis}}
  \emph{(\bibinfo{series}{ISSTA '22})}. \bibinfo{pages}{340--351}.
\newblock
\urldef\tempurl%
\url{https://doi.org/10.1145/3533767.3534376}
\showDOI{\tempurl}


\bibitem[\protect\citeauthoryear{Aschermann, Frassetto, Holz, Jauernig,
  Sadeghi, and Teuchert}{Aschermann et~al\mbox{.}}{2019a}]%
        {nautilus2019}
\bibfield{author}{\bibinfo{person}{Cornelius Aschermann},
  \bibinfo{person}{Tommaso Frassetto}, \bibinfo{person}{Thorsten Holz},
  \bibinfo{person}{Patrick Jauernig}, \bibinfo{person}{Ahmad-Reza Sadeghi},
  {and} \bibinfo{person}{Daniel Teuchert}.} \bibinfo{year}{2019}\natexlab{a}.
\newblock \showarticletitle{NAUTILUS: Fishing for Deep Bugs with Grammars}. In
  \bibinfo{booktitle}{\emph{Network and Distributed Systems Security
  Symposium}} \emph{(\bibinfo{series}{NDSS '19})}.
\newblock
\urldef\tempurl%
\url{https://doi.org/10.14722/ndss.2019.23412}
\showDOI{\tempurl}


\bibitem[\protect\citeauthoryear{Aschermann, Schumilo, Blazytko, Gawlik, and
  Holz}{Aschermann et~al\mbox{.}}{2019b}]%
        {redqueen2019}
\bibfield{author}{\bibinfo{person}{Cornelius Aschermann},
  \bibinfo{person}{Sergej Schumilo}, \bibinfo{person}{Tim Blazytko},
  \bibinfo{person}{Robert Gawlik}, {and} \bibinfo{person}{Thorsten Holz}.}
  \bibinfo{year}{2019}\natexlab{b}.
\newblock \showarticletitle{REDQUEEN: Fuzzing with Input-to-State
  Correspondence}. In \bibinfo{booktitle}{\emph{Symposium on Network and
  Distributed System Security}} \emph{(\bibinfo{series}{NDSS '19})}.
  \bibinfo{pages}{1--15}.
\newblock
\urldef\tempurl%
\url{https://doi.org/10.14722/ndss.2019.23371}
\showDOI{\tempurl}


\bibitem[\protect\citeauthoryear{Biondo}{Biondo}{2018}]%
        {improvingaflqemu2018}
\bibfield{author}{\bibinfo{person}{Andrea Biondo}.}
  \bibinfo{year}{2018}\natexlab{}.
\newblock \bibinfo{title}{Improving {AFL}'s QEMU mode performance}.
\newblock
  \bibinfo{howpublished}{\url{https://abiondo.me/2018/09/21/improving-afl-qemu-mode}}.
\newblock


\bibitem[\protect\citeauthoryear{Blazytko, Bishop, Aschermann, Cappos,
  Schl{\"o}gel, Korshun, Abbasi, Schweighauser, Schinzel, Schumilo,
  et~al\mbox{.}}{Blazytko et~al\mbox{.}}{2019}]%
        {grimoire2019}
\bibfield{author}{\bibinfo{person}{Tim Blazytko}, \bibinfo{person}{Matt
  Bishop}, \bibinfo{person}{Cornelius Aschermann}, \bibinfo{person}{Justin
  Cappos}, \bibinfo{person}{Moritz Schl{\"o}gel}, \bibinfo{person}{Nadia
  Korshun}, \bibinfo{person}{Ali Abbasi}, \bibinfo{person}{Marco
  Schweighauser}, \bibinfo{person}{Sebastian Schinzel}, \bibinfo{person}{Sergej
  Schumilo}, {et~al\mbox{.}}} \bibinfo{year}{2019}\natexlab{}.
\newblock \showarticletitle{{GRIMOIRE}: Synthesizing Structure while Fuzzing}.
  In \bibinfo{booktitle}{\emph{28th {USENIX} Security Symposium}}
  \emph{(\bibinfo{series}{USENIX Security '19})}. \bibinfo{pages}{1985--2002}.
\newblock
\urldef\tempurl%
\url{https://doi.org/10.5555/3361338.3361475}
\showDOI{\tempurl}


\bibitem[\protect\citeauthoryear{B{\"o}hme, Szekeres, and Metzman}{B{\"o}hme
  et~al\mbox{.}}{2022}]%
        {covbenchreliability2022}
\bibfield{author}{\bibinfo{person}{Marcel B{\"o}hme},
  \bibinfo{person}{L{\'a}szl{\'o} Szekeres}, {and} \bibinfo{person}{Jonathan
  Metzman}.} \bibinfo{year}{2022}\natexlab{}.
\newblock \showarticletitle{On the Reliability of Coverage-Based Fuzzer
  Benchmarking}. In \bibinfo{booktitle}{\emph{44th IEEE/ACM International
  Conference on Software Engineering}} \emph{(\bibinfo{series}{ICSE '22})}.
\newblock
\urldef\tempurl%
\url{https://doi.org/10.1145/3510003.3510230}
\showDOI{\tempurl}


\bibitem[\protect\citeauthoryear{Bruening, Zhao, and Amarasinghe}{Bruening
  et~al\mbox{.}}{2012}]%
        {dynamorio2012}
\bibfield{author}{\bibinfo{person}{Derek Bruening}, \bibinfo{person}{Qin Zhao},
  {and} \bibinfo{person}{Saman Amarasinghe}.} \bibinfo{year}{2012}\natexlab{}.
\newblock \showarticletitle{Transparent dynamic instrumentation}. In
  \bibinfo{booktitle}{\emph{Proceedings of the 8th ACM SIGPLAN/SIGOPS
  conference on Virtual Execution Environments}} \emph{(\bibinfo{series}{VEE
  '23})}. \bibinfo{pages}{133--144}.
\newblock
\urldef\tempurl%
\url{https://doi.org/10.1145/2151024.2151043}
\showDOI{\tempurl}


\bibitem[\protect\citeauthoryear{Brumley, Jager, Avgerinos, and
  Schwartz}{Brumley et~al\mbox{.}}{2011}]%
        {bap2011}
\bibfield{author}{\bibinfo{person}{David Brumley}, \bibinfo{person}{Ivan
  Jager}, \bibinfo{person}{Thanassis Avgerinos}, {and}
  \bibinfo{person}{Edward~J Schwartz}.} \bibinfo{year}{2011}\natexlab{}.
\newblock \showarticletitle{BAP: A binary analysis platform}. In
  \bibinfo{booktitle}{\emph{Proceedings of the 23rd International Conference on
  Computer Aided Verification}} \emph{(\bibinfo{series}{CAV '11})}.
  \bibinfo{pages}{463--469}.
\newblock
\urldef\tempurl%
\url{https://doi.org/10.5555/2032305.2032342}
\showDOI{\tempurl}


\bibitem[\protect\citeauthoryear{Chen, Woo, Brumley, and Egele}{Chen
  et~al\mbox{.}}{2016}]%
        {firmadyne2016}
\bibfield{author}{\bibinfo{person}{Daming~D Chen}, \bibinfo{person}{Maverick
  Woo}, \bibinfo{person}{David Brumley}, {and} \bibinfo{person}{Manuel Egele}.}
  \bibinfo{year}{2016}\natexlab{}.
\newblock \showarticletitle{Towards automated dynamic analysis for linux-based
  embedded firmware}. In \bibinfo{booktitle}{\emph{Network and Distributed
  System Security Symposium}} \emph{(\bibinfo{series}{NDSS})}.
\newblock
\urldef\tempurl%
\url{https://doi.org/10.14722/ndss.2016.23415}
\showDOI{\tempurl}


\bibitem[\protect\citeauthoryear{Chen and Chen}{Chen and Chen}{2018}]%
        {angora2018}
\bibfield{author}{\bibinfo{person}{Peng Chen} {and} \bibinfo{person}{Hao
  Chen}.} \bibinfo{year}{2018}\natexlab{}.
\newblock \showarticletitle{Angora: Efficient fuzzing by principled search}. In
  \bibinfo{booktitle}{\emph{{IEEE} Symposium on Security and Privacy}}
  \emph{(\bibinfo{series}{SP})}. \bibinfo{pages}{711--725}.
\newblock
\urldef\tempurl%
\url{https://doi.org/10.1109/SP.2018.00046}
\showDOI{\tempurl}


\bibitem[\protect\citeauthoryear{Chen, Thomas, and Garcia}{Chen
  et~al\mbox{.}}{2022}]%
        {metaemu2022}
\bibfield{author}{\bibinfo{person}{Zitai Chen}, \bibinfo{person}{Sam~L Thomas},
  {and} \bibinfo{person}{Flavio~D Garcia}.} \bibinfo{year}{2022}\natexlab{}.
\newblock \showarticletitle{MetaEmu: An Architecture Agnostic Rehosting
  Framework for Automotive Firmware}. In \bibinfo{booktitle}{\emph{Proceedings
  of the 2022 ACM SIGSAC Conference on Computer and Communications Security}}
  \emph{(\bibinfo{series}{CCS '22})}. \bibinfo{pages}{515--529}.
\newblock
\urldef\tempurl%
\url{https://doi.org/10.1145/3548606.3559338}
\showDOI{\tempurl}


\bibitem[\protect\citeauthoryear{Chipounov, Kuznetsov, and Candea}{Chipounov
  et~al\mbox{.}}{2011}]%
        {s2e2011}
\bibfield{author}{\bibinfo{person}{Vitaly Chipounov},
  \bibinfo{person}{Volodymyr Kuznetsov}, {and} \bibinfo{person}{George
  Candea}.} \bibinfo{year}{2011}\natexlab{}.
\newblock \showarticletitle{S2E: A platform for in-vivo multi-path analysis of
  software systems}. In \bibinfo{booktitle}{\emph{Proceedings of the 16th
  International Conference on Architectural Support for Programming Languages
  and Operating Systems}} \emph{(\bibinfo{series}{ASPLOS XVI})}.
  \bibinfo{pages}{265--278}.
\newblock
\urldef\tempurl%
\url{https://doi.org/10.1145/1950365.1950396}
\showDOI{\tempurl}


\bibitem[\protect\citeauthoryear{Clements, Gustafson, Scharnowski, Grosen,
  Fritz, Kruegel, Vigna, Bagchi, and Payer}{Clements et~al\mbox{.}}{2020}]%
        {halucinator2020}
\bibfield{author}{\bibinfo{person}{Abraham~A Clements}, \bibinfo{person}{Eric
  Gustafson}, \bibinfo{person}{Tobias Scharnowski}, \bibinfo{person}{Paul
  Grosen}, \bibinfo{person}{David Fritz}, \bibinfo{person}{Christopher
  Kruegel}, \bibinfo{person}{Giovanni Vigna}, \bibinfo{person}{Saurabh Bagchi},
  {and} \bibinfo{person}{Mathias Payer}.} \bibinfo{year}{2020}\natexlab{}.
\newblock \showarticletitle{{HALucinator}: Firmware Re-hosting Through
  Abstraction Layer Emulation}. In \bibinfo{booktitle}{\emph{29th {USENIX}
  Security Symposium}} \emph{(\bibinfo{series}{USENIX Security '20})}.
  \bibinfo{pages}{1201--1218}.
\newblock
\showISBNx{978-1-939133-17-5}


\bibitem[\protect\citeauthoryear{Contributors}{Contributors}{2022}]%
        {qlinggithub2022}
\bibfield{author}{\bibinfo{person}{The~Qiling Contributors}.}
  \bibinfo{year}{2022}\natexlab{}.
\newblock \bibinfo{title}{Qiling Advanced Binary Emulation Framework}.
\newblock
\newblock
\urldef\tempurl%
\url{https://github.com/qilingframework/qiling/}
\showURL{%
\tempurl}


\bibitem[\protect\citeauthoryear{Corteggiani, Camurati, and
  Francillon}{Corteggiani et~al\mbox{.}}{2018}]%
        {inception2018}
\bibfield{author}{\bibinfo{person}{Nassim Corteggiani},
  \bibinfo{person}{Giovanni Camurati}, {and} \bibinfo{person}{Aur{\'e}lien
  Francillon}.} \bibinfo{year}{2018}\natexlab{}.
\newblock \showarticletitle{Inception: System-Wide Security Testing of
  Real-World Embedded Systems Software}. In \bibinfo{booktitle}{\emph{27th
  {USENIX} Security Symposium}} \emph{(\bibinfo{series}{USENIX Security '18})}.
  \bibinfo{pages}{309--326}.
\newblock
\showISBNx{978-1-939133-04-5}


\bibitem[\protect\citeauthoryear{Davidson, Moench, Ristenpart, and
  Jha}{Davidson et~al\mbox{.}}{2013}]%
        {fie2013}
\bibfield{author}{\bibinfo{person}{Drew Davidson}, \bibinfo{person}{Benjamin
  Moench}, \bibinfo{person}{Thomas Ristenpart}, {and} \bibinfo{person}{Somesh
  Jha}.} \bibinfo{year}{2013}\natexlab{}.
\newblock \showarticletitle{{FiE} on Firmware: Finding Vulnerabilities in
  Embedded Systems Using Symbolic Execution}. In \bibinfo{booktitle}{\emph{22nd
  {USENIX} Security Symposium}} \emph{(\bibinfo{series}{USENIX Security '13})}.
  \bibinfo{pages}{463--478}.
\newblock
\urldef\tempurl%
\url{https://doi.org/10.5555/2534766.2534806}
\showDOI{\tempurl}


\bibitem[\protect\citeauthoryear{Dinesh, Burow, Xu, and Payer}{Dinesh
  et~al\mbox{.}}{2020}]%
        {retrowrite2020}
\bibfield{author}{\bibinfo{person}{Sushant Dinesh}, \bibinfo{person}{Nathan
  Burow}, \bibinfo{person}{Dongyan Xu}, {and} \bibinfo{person}{Mathias Payer}.}
  \bibinfo{year}{2020}\natexlab{}.
\newblock \showarticletitle{RetroWrite: Statically Instrumenting COTS Binaries
  for Fuzzing and Sanitization}. In \bibinfo{booktitle}{\emph{{IEEE} Symposium
  on Security and Privacy}} \emph{(\bibinfo{series}{SP '20})}.
  \bibinfo{pages}{1497--1511}.
\newblock
\urldef\tempurl%
\url{https://doi.org/10.1109/SP40000.2020.00009}
\showDOI{\tempurl}


\bibitem[\protect\citeauthoryear{Dola}{Dola}{2021}]%
        {aflghidraemu2021}
\bibfield{author}{\bibinfo{person}{Flavian Dola}.}
  \bibinfo{year}{2021}\natexlab{}.
\newblock \bibinfo{title}{Fuzzing exotic arch with AFL using ghidra emulator}.
\newblock
  \bibinfo{howpublished}{\url{https://airbus-cyber-security.com/fuzzing-exotic-arch-with-afl-using-ghidra-emulator/}}.
\newblock


\bibitem[\protect\citeauthoryear{Dolan-Gavitt, Hulin, Kirda, Leek, Mambretti,
  Robertson, Ulrich, and Whelan}{Dolan-Gavitt et~al\mbox{.}}{2016}]%
        {lava2016}
\bibfield{author}{\bibinfo{person}{Brendan Dolan-Gavitt},
  \bibinfo{person}{Patrick Hulin}, \bibinfo{person}{Engin Kirda},
  \bibinfo{person}{Tim Leek}, \bibinfo{person}{Andrea Mambretti},
  \bibinfo{person}{Wil Robertson}, \bibinfo{person}{Frederick Ulrich}, {and}
  \bibinfo{person}{Ryan Whelan}.} \bibinfo{year}{2016}\natexlab{}.
\newblock \showarticletitle{LAVA: Large-scale Automated Vulnerability
  Addition}. In \bibinfo{booktitle}{\emph{{IEEE} Symposium on Security and
  Privacy}} \emph{(\bibinfo{series}{SP '16})}. \bibinfo{pages}{110--121}.
\newblock
\urldef\tempurl%
\url{https://doi.org/10.1109/SP.2016.15}
\showDOI{\tempurl}


\bibitem[\protect\citeauthoryear{Falk}{Falk}{2018}]%
        {softserve2018}
\bibfield{author}{\bibinfo{person}{Brandon Falk}.}
  \bibinfo{year}{2018}\natexlab{}.
\newblock \bibinfo{title}{Vectorized Emulation: MMU Design}.
\newblock
  \bibinfo{howpublished}{\url{https://gamozolabs.github.io/fuzzing/2018/11/19/vectorized_emulation_mmu.html}}.
\newblock


\bibitem[\protect\citeauthoryear{Farrelly, Chesser, and Ranasinghe}{Farrelly
  et~al\mbox{.}}{2023}]%
        {emberio2023}
\bibfield{author}{\bibinfo{person}{Guy Farrelly}, \bibinfo{person}{Michael
  Chesser}, {and} \bibinfo{person}{Damith~C. Ranasinghe}.}
  \bibinfo{year}{2023}\natexlab{}.
\newblock \showarticletitle{{Ember-IO}: Effective Firmware Fuzzing with
  Model-Free Memory Mapped IO}. In \bibinfo{booktitle}{\emph{Proceedings of
  18th ACM ASIA Conference on Computer and Communications Security}}
  \emph{(\bibinfo{series}{AsiaCCS '23})}.
\newblock
\urldef\tempurl%
\url{https://doi.org/10.1145/3579856.3582840}
\showDOI{\tempurl}


\bibitem[\protect\citeauthoryear{Feng, Mera, and Lu}{Feng
  et~al\mbox{.}}{2020}]%
        {p2im2020}
\bibfield{author}{\bibinfo{person}{Bo Feng}, \bibinfo{person}{Alejandro Mera},
  {and} \bibinfo{person}{Long Lu}.} \bibinfo{year}{2020}\natexlab{}.
\newblock \showarticletitle{{P$^2$IM}: Scalable and Hardware-independent
  Firmware Testing via Automatic Peripheral Interface Modeling}. In
  \bibinfo{booktitle}{\emph{29th {USENIX} Security Symposium}}
  \emph{(\bibinfo{series}{USENIX Security '20})}. \bibinfo{pages}{1237--1254}.
\newblock
\showISBNx{978-1-939133-17-5}


\bibitem[\protect\citeauthoryear{Fioraldi, D'Elia, and Coppa}{Fioraldi
  et~al\mbox{.}}{2020a}]%
        {weizz2020}
\bibfield{author}{\bibinfo{person}{Andrea Fioraldi},
  \bibinfo{person}{Daniele~Cono D'Elia}, {and} \bibinfo{person}{Emilio Coppa}.}
  \bibinfo{year}{2020}\natexlab{a}.
\newblock \showarticletitle{WEIZZ: Automatic grey-box fuzzing for structured
  binary formats}. In \bibinfo{booktitle}{\emph{Proceedings of the 29th {ACM
  SIGSOFT} International Symposium on Software Testing and Analysis}}
  \emph{(\bibinfo{series}{ISSTA '20})}. \bibinfo{pages}{1--13}.
\newblock
\urldef\tempurl%
\url{https://doi.org/10.1145/3395363.3397372}
\showDOI{\tempurl}


\bibitem[\protect\citeauthoryear{Fioraldi, D'Elia, and Querzoni}{Fioraldi
  et~al\mbox{.}}{2020b}]%
        {qasan2020}
\bibfield{author}{\bibinfo{person}{Andrea Fioraldi},
  \bibinfo{person}{Daniele~Cono D'Elia}, {and} \bibinfo{person}{Leonardo
  Querzoni}.} \bibinfo{year}{2020}\natexlab{b}.
\newblock \showarticletitle{Fuzzing Binaries for Memory Safety Errors with
  {QASan}}. In \bibinfo{booktitle}{\emph{{IEEE} Secure Development Conference}}
  \emph{(\bibinfo{series}{SecDev '20})}. \bibinfo{pages}{23--30}.
\newblock
\urldef\tempurl%
\url{https://doi.org/10.1109/SecDev45635.2020.00019}
\showDOI{\tempurl}


\bibitem[\protect\citeauthoryear{Fioraldi, Maier, Ei{\ss}feldt, and
  Heuse}{Fioraldi et~al\mbox{.}}{2020c}]%
        {aflplusplus2020}
\bibfield{author}{\bibinfo{person}{Andrea Fioraldi}, \bibinfo{person}{Dominik
  Maier}, \bibinfo{person}{Heiko Ei{\ss}feldt}, {and} \bibinfo{person}{Marc
  Heuse}.} \bibinfo{year}{2020}\natexlab{c}.
\newblock \showarticletitle{AFL++: Combining Incremental Steps of Fuzzing
  Research}. In \bibinfo{booktitle}{\emph{14th {USENIX} Workshop on Offensive
  Technologies}} \emph{(\bibinfo{series}{WOOT '20})}.
\newblock
\urldef\tempurl%
\url{https://doi.org/10.5555/3488877.3488887}
\showDOI{\tempurl}


\bibitem[\protect\citeauthoryear{Flores-Montoya and Schulte}{Flores-Montoya and
  Schulte}{2020}]%
        {datalogdisasm2020}
\bibfield{author}{\bibinfo{person}{Antonio Flores-Montoya} {and}
  \bibinfo{person}{Eric Schulte}.} \bibinfo{year}{2020}\natexlab{}.
\newblock \showarticletitle{Datalog Disassembly}. In
  \bibinfo{booktitle}{\emph{29th {USENIX} Security Symposium}}
  \emph{(\bibinfo{series}{USENIX Security '20})}. \bibinfo{pages}{1075--1092}.
\newblock
\urldef\tempurl%
\url{https://doi.org/10.5555/3489212.3489273}
\showDOI{\tempurl}


\bibitem[\protect\citeauthoryear{Foundation}{Foundation}{2022}]%
        {gnucoreutils}
\bibfield{author}{\bibinfo{person}{Free~Software Foundation}.}
  \bibinfo{year}{2022}\natexlab{}.
\newblock \bibinfo{title}{GNU core utilities}.
\newblock
  \bibinfo{howpublished}{\url{https://www.gnu.org/software/coreutils/}}.
\newblock


\bibitem[\protect\citeauthoryear{Gan, Zhang, Chen, Zhao, Qin, Wu, and Chen}{Gan
  et~al\mbox{.}}{2020}]%
        {greyone2020}
\bibfield{author}{\bibinfo{person}{Shuitao Gan}, \bibinfo{person}{Chao Zhang},
  \bibinfo{person}{Peng Chen}, \bibinfo{person}{Bodong Zhao},
  \bibinfo{person}{Xiaojun Qin}, \bibinfo{person}{Dong Wu}, {and}
  \bibinfo{person}{Zuoning Chen}.} \bibinfo{year}{2020}\natexlab{}.
\newblock \showarticletitle{GREYONE: Data Flow Sensitive Fuzzing}. In
  \bibinfo{booktitle}{\emph{29th USENIX Security Symposium (USENIX Security
  20)}}. \bibinfo{pages}{2577--2594}.
\newblock


\bibitem[\protect\citeauthoryear{Gui, Shu, Kang, and Xiong}{Gui
  et~al\mbox{.}}{2020}]%
        {firmcorn2020}
\bibfield{author}{\bibinfo{person}{Zhijie Gui}, \bibinfo{person}{Hui Shu},
  \bibinfo{person}{Fei Kang}, {and} \bibinfo{person}{Xiaobing Xiong}.}
  \bibinfo{year}{2020}\natexlab{}.
\newblock \showarticletitle{{FIRMCORN}: Vulnerability-Oriented Fuzzing of {IoT}
  Firmware via Optimized Virtual Execution}.
\newblock \bibinfo{journal}{\emph{IEEE Access}}  \bibinfo{volume}{8}
  (\bibinfo{year}{2020}), \bibinfo{pages}{29826--29841}.
\newblock
\urldef\tempurl%
\url{https://doi.org/10.1109/ACCESS.2020.2973043}
\showDOI{\tempurl}


\bibitem[\protect\citeauthoryear{Gustafson, Muench, Spensky, Redini, Machiry,
  Fratantonio, Balzarotti, Francillon, Choe, Kruegel, and Vigna}{Gustafson
  et~al\mbox{.}}{2019}]%
        {pretender2019}
\bibfield{author}{\bibinfo{person}{Eric Gustafson}, \bibinfo{person}{Marius
  Muench}, \bibinfo{person}{Chad Spensky}, \bibinfo{person}{Nilo Redini},
  \bibinfo{person}{Aravind Machiry}, \bibinfo{person}{Yanick Fratantonio},
  \bibinfo{person}{Davide Balzarotti}, \bibinfo{person}{Aur{\'e}lien
  Francillon}, \bibinfo{person}{Yung~Ryn Choe}, \bibinfo{person}{Christophe
  Kruegel}, {and} \bibinfo{person}{Giovanni Vigna}.}
  \bibinfo{year}{2019}\natexlab{}.
\newblock \showarticletitle{Toward the Analysis of Embedded Firmware through
  Automated Re-hosting}. In \bibinfo{booktitle}{\emph{22nd International
  Symposium on Research in Attacks, Intrusions and Defenses}}
  \emph{(\bibinfo{series}{RAID '19})}. \bibinfo{pages}{135--150}.
\newblock
\showISBNx{978-1-939133-07-6}


\bibitem[\protect\citeauthoryear{Hazimeh, Herrera, and Payer}{Hazimeh
  et~al\mbox{.}}{2020}]%
        {magma2020}
\bibfield{author}{\bibinfo{person}{Ahmad Hazimeh}, \bibinfo{person}{Adrian
  Herrera}, {and} \bibinfo{person}{Mathias Payer}.}
  \bibinfo{year}{2020}\natexlab{}.
\newblock \showarticletitle{Magma: A Ground-Truth Fuzzing Benchmark}.
\newblock \bibinfo{journal}{\emph{Proceedings of the ACM on Measurement and
  Analysis of Computing Systems}} \bibinfo{volume}{4}, \bibinfo{number}{3}
  (\bibinfo{year}{2020}).
\newblock
\urldef\tempurl%
\url{https://doi.org/10.1145/3428334}
\showDOI{\tempurl}


\bibitem[\protect\citeauthoryear{Herdt, Gro{\ss}e, Le, and Drechsler}{Herdt
  et~al\mbox{.}}{2019}]%
        {verifyingiss2019}
\bibfield{author}{\bibinfo{person}{Vladimir Herdt}, \bibinfo{person}{Daniel
  Gro{\ss}e}, \bibinfo{person}{Hoang~M Le}, {and} \bibinfo{person}{Rolf
  Drechsler}.} \bibinfo{year}{2019}\natexlab{}.
\newblock \showarticletitle{Verifying Instruction Set Simulators using
  Coverage-guided Fuzzing}. In \bibinfo{booktitle}{\emph{Design, Automation \&
  Test in Europe Conference \& Exhibition}} \emph{(\bibinfo{series}{DATE'
  19})}. \bibinfo{pages}{360--365}.
\newblock
\urldef\tempurl%
\url{https://doi.org/10.23919/DATE.2019.8714912}
\showDOI{\tempurl}


\bibitem[\protect\citeauthoryear{Intel}{Intel}{2016}]%
        {lafintel2016}
\bibfield{author}{\bibinfo{person}{Intel}.} \bibinfo{year}{2016}\natexlab{}.
\newblock \bibinfo{title}{Circumventing fuzzing roadblocks with compiler
  transformations}.
\newblock
  \bibinfo{howpublished}{\url{https://lafintel.wordpress.com/2016/08/15/circumventing-fuzzing-roadblocks-with-compiler-transformations/}}.
\newblock


\bibitem[\protect\citeauthoryear{Kammerstetter, Burian, and
  Kastner}{Kammerstetter et~al\mbox{.}}{2016}]%
        {embedded2016}
\bibfield{author}{\bibinfo{person}{Markus Kammerstetter},
  \bibinfo{person}{Daniel Burian}, {and} \bibinfo{person}{Wolfgang Kastner}.}
  \bibinfo{year}{2016}\natexlab{}.
\newblock \showarticletitle{Embedded Security Testing with Peripheral Device
  Caching and Runtime Program State Approximation}. In
  \bibinfo{booktitle}{\emph{10th International Conference on Emerging Security
  Information, Systems and Technologies}} \emph{(\bibinfo{series}{SECUWARE
  '16})}.
\newblock


\bibitem[\protect\citeauthoryear{Klees, Ruef, Cooper, Wei, and Hicks}{Klees
  et~al\mbox{.}}{2018}]%
        {fuzzeval2018}
\bibfield{author}{\bibinfo{person}{George~T. Klees}, \bibinfo{person}{Andrew
  Ruef}, \bibinfo{person}{Benjamin Cooper}, \bibinfo{person}{Shiyi Wei}, {and}
  \bibinfo{person}{Michael Hicks}.} \bibinfo{year}{2018}\natexlab{}.
\newblock \showarticletitle{Evaluating Fuzz Testing}. In
  \bibinfo{booktitle}{\emph{Proceedings of the {ACM} Conference on Computer and
  Communications Security}} \emph{(\bibinfo{series}{CCS '18})}.
  \bibinfo{pages}{2123--2138}.
\newblock
\urldef\tempurl%
\url{https://doi.org/10.1145/3243734.3243804}
\showDOI{\tempurl}


\bibitem[\protect\citeauthoryear{Laboratory}{Laboratory}{2016}]%
        {msp430qemu2016}
\bibfield{author}{\bibinfo{person}{Draper Laboratory}.}
  \bibinfo{year}{2016}\natexlab{}.
\newblock \bibinfo{title}{QEMU MSP430 Target}.
\newblock
  \bibinfo{howpublished}{\url{https://github.com/draperlaboratory/qemu-msp}}.
\newblock


\bibitem[\protect\citeauthoryear{Luk, Cohn, Muth, Patil, Klauser, Lowney,
  Wallace, Reddi, and Hazelwood}{Luk et~al\mbox{.}}{2005}]%
        {pin2005}
\bibfield{author}{\bibinfo{person}{Chi-Keung Luk}, \bibinfo{person}{Robert
  Cohn}, \bibinfo{person}{Robert Muth}, \bibinfo{person}{Harish Patil},
  \bibinfo{person}{Artur Klauser}, \bibinfo{person}{Geoff Lowney},
  \bibinfo{person}{Steven Wallace}, \bibinfo{person}{Vijay~Janapa Reddi}, {and}
  \bibinfo{person}{Kim Hazelwood}.} \bibinfo{year}{2005}\natexlab{}.
\newblock \showarticletitle{Pin: Building Customized Program Analysis Tools
  with Dynamic Instrumentation}. In \bibinfo{booktitle}{\emph{Proceedings of
  the 2005 ACM SIGPLAN Conference on Programming Language Design and
  Implementation}} \emph{(\bibinfo{series}{PLDI '05})}.
  \bibinfo{pages}{190--200}.
\newblock
\urldef\tempurl%
\url{https://doi.org/10.1145/1064978.1065034}
\showDOI{\tempurl}


\bibitem[\protect\citeauthoryear{Ma, Zhao, Ren, Li, Ma, Luo, and Zhang}{Ma
  et~al\mbox{.}}{2022}]%
        {printfuzz2022}
\bibfield{author}{\bibinfo{person}{Zheyu Ma}, \bibinfo{person}{Bodong Zhao},
  \bibinfo{person}{Letu Ren}, \bibinfo{person}{Zheming Li},
  \bibinfo{person}{Siqi Ma}, \bibinfo{person}{Xiapu Luo}, {and}
  \bibinfo{person}{Chao Zhang}.} \bibinfo{year}{2022}\natexlab{}.
\newblock \showarticletitle{PrIntFuzz: fuzzing Linux drivers via automated
  virtual device simulation}. In \bibinfo{booktitle}{\emph{Proceedings of the
  31st ACM SIGSOFT International Symposium on Software Testing and Analysis}}
  \emph{(\bibinfo{series}{ISSTA '22})}. \bibinfo{pages}{404--416}.
\newblock
\urldef\tempurl%
\url{https://doi.org/10.1145/3533767.3534226}
\showDOI{\tempurl}


\bibitem[\protect\citeauthoryear{Maier, Radtke, and Harren}{Maier
  et~al\mbox{.}}{2019}]%
        {unicorefuzz2019}
\bibfield{author}{\bibinfo{person}{Dominik Maier}, \bibinfo{person}{Benedikt
  Radtke}, {and} \bibinfo{person}{Bastian Harren}.}
  \bibinfo{year}{2019}\natexlab{}.
\newblock \showarticletitle{Unicorefuzz: On the Viability of Emulation for
  Kernelspace Fuzzing}. In \bibinfo{booktitle}{\emph{13th {USENIX} Workshop on
  Offensive Technologies}} \emph{(\bibinfo{series}{WOOT '19})}.
\newblock
\urldef\tempurl%
\url{https://doi.org/10.5555/3359043.3359051}
\showDOI{\tempurl}


\bibitem[\protect\citeauthoryear{Maier, Seidel, and Park}{Maier
  et~al\mbox{.}}{2020}]%
        {basesafe2020}
\bibfield{author}{\bibinfo{person}{Dominik Maier}, \bibinfo{person}{Lukas
  Seidel}, {and} \bibinfo{person}{Shinjo Park}.}
  \bibinfo{year}{2020}\natexlab{}.
\newblock \showarticletitle{BaseSAFE: Baseband SAnitized Fuzzing through
  Emulation}. In \bibinfo{booktitle}{\emph{Proceedings of the 13th ACM
  Conference on Security and Privacy in Wireless and Mobile Networks}}
  \emph{(\bibinfo{series}{WiSec '20})}. \bibinfo{pages}{122--132}.
\newblock
\urldef\tempurl%
\url{https://doi.org/10.1145/3395351.3399360}
\showDOI{\tempurl}


\bibitem[\protect\citeauthoryear{Martignoni, McCamant, Poosankam, Song, and
  Maniatis}{Martignoni et~al\mbox{.}}{2012}]%
        {pokeemu2012}
\bibfield{author}{\bibinfo{person}{Lorenzo Martignoni},
  \bibinfo{person}{Stephen McCamant}, \bibinfo{person}{Pongsin Poosankam},
  \bibinfo{person}{Dawn Song}, {and} \bibinfo{person}{Petros Maniatis}.}
  \bibinfo{year}{2012}\natexlab{}.
\newblock \showarticletitle{Path-Exploration Lifting: Hi-Fi Tests for Lo-Fi
  Emulators}. In \bibinfo{booktitle}{\emph{Proceedings of the Seventeenth
  International Conference on Architectural Support for Programming Languages
  and Operating Systems}} \emph{(\bibinfo{series}{ASPLOS XVII})}.
  \bibinfo{pages}{337--348}.
\newblock
\urldef\tempurl%
\url{https://doi.org/10.1145/2150976.2151012}
\showDOI{\tempurl}


\bibitem[\protect\citeauthoryear{Martignoni, Paleari, Fresi~Roglia, and
  Bruschi}{Martignoni et~al\mbox{.}}{2010}]%
        {kemufuzz2010}
\bibfield{author}{\bibinfo{person}{Lorenzo Martignoni},
  \bibinfo{person}{Roberto Paleari}, \bibinfo{person}{Giampaolo Fresi~Roglia},
  {and} \bibinfo{person}{Danilo Bruschi}.} \bibinfo{year}{2010}\natexlab{}.
\newblock \showarticletitle{Testing System Virtual Machines}. In
  \bibinfo{booktitle}{\emph{Proceedings of the 19th International Symposium on
  Software Testing and Analysis}} \emph{(\bibinfo{series}{ISSTA '10})}.
  \bibinfo{pages}{171–182}.
\newblock
\showISBNx{9781605588230}
\urldef\tempurl%
\url{https://doi.org/10.1145/1831708.1831730}
\showDOI{\tempurl}


\bibitem[\protect\citeauthoryear{Martignoni, Paleari, Roglia, and
  Bruschi}{Martignoni et~al\mbox{.}}{2009}]%
        {emufuzzer2009}
\bibfield{author}{\bibinfo{person}{Lorenzo Martignoni},
  \bibinfo{person}{Roberto Paleari}, \bibinfo{person}{Giampaolo~Fresi Roglia},
  {and} \bibinfo{person}{Danilo Bruschi}.} \bibinfo{year}{2009}\natexlab{}.
\newblock \showarticletitle{Testing CPU emulators}. In
  \bibinfo{booktitle}{\emph{Proceedings of the 18th International Symposium on
  Software Testing and Analysis}} \emph{(\bibinfo{series}{ISSTA' 09})}.
  \bibinfo{pages}{261--272}.
\newblock
\urldef\tempurl%
\url{https://doi.org/10.1145/1572272.1572303}
\showDOI{\tempurl}


\bibitem[\protect\citeauthoryear{Metzman, Szekeres, Simon, Sprabery, and
  Arya}{Metzman et~al\mbox{.}}{2021}]%
        {fuzzbench2021}
\bibfield{author}{\bibinfo{person}{Jonathan Metzman},
  \bibinfo{person}{L{\'a}szl{\'o} Szekeres}, \bibinfo{person}{Laurent Simon},
  \bibinfo{person}{Read Sprabery}, {and} \bibinfo{person}{Abhishek Arya}.}
  \bibinfo{year}{2021}\natexlab{}.
\newblock \showarticletitle{FuzzBench: an open fuzzer benchmarking platform and
  service}. In \bibinfo{booktitle}{\emph{Proceedings of the 29th ACM Joint
  Meeting on European Software Engineering Conference and Symposium on the
  Foundations of Software Engineering}} \emph{(\bibinfo{series}{ESEC/FSE
  '21})}. \bibinfo{pages}{1393--1403}.
\newblock
\urldef\tempurl%
\url{https://doi.org/10.1145/3468264.3473932}
\showDOI{\tempurl}


\bibitem[\protect\citeauthoryear{Muench, Nisi, Francillon, and
  Balzarotti}{Muench et~al\mbox{.}}{2018}]%
        {avatar2018}
\bibfield{author}{\bibinfo{person}{Marius Muench}, \bibinfo{person}{Dario
  Nisi}, \bibinfo{person}{Aur{\'e}lien Francillon}, {and}
  \bibinfo{person}{Davide Balzarotti}.} \bibinfo{year}{2018}\natexlab{}.
\newblock \showarticletitle{Avatar$^2$: A multi-target orchestration platform}.
  In \bibinfo{booktitle}{\emph{Workshop on Binary Analysis Research}}
  \emph{(\bibinfo{series}{BAR '18})}.
\newblock
\urldef\tempurl%
\url{https://doi.org/10.14722/bar.2018.23017}
\showDOI{\tempurl}


\bibitem[\protect\citeauthoryear{Nagy, Nguyen-Tuong, Hiser, Davidson, and
  Hicks}{Nagy et~al\mbox{.}}{2021}]%
        {zafl2021}
\bibfield{author}{\bibinfo{person}{Stefan Nagy}, \bibinfo{person}{Anh
  Nguyen-Tuong}, \bibinfo{person}{Jason~D Hiser}, \bibinfo{person}{Jack~W
  Davidson}, {and} \bibinfo{person}{Matthew Hicks}.}
  \bibinfo{year}{2021}\natexlab{}.
\newblock \showarticletitle{Breaking Through Binaries: Compiler-quality
  Instrumentation for Better Binary-only Fuzzing}. In
  \bibinfo{booktitle}{\emph{30th {USENIX} Security Symposium}}
  \emph{(\bibinfo{series}{USENIX Security '21})}. \bibinfo{pages}{1683--1700}.
\newblock


\bibitem[\protect\citeauthoryear{Nethercote and Seward}{Nethercote and
  Seward}{2007}]%
        {valgrind2007}
\bibfield{author}{\bibinfo{person}{Nicholas Nethercote} {and}
  \bibinfo{person}{Julian Seward}.} \bibinfo{year}{2007}\natexlab{}.
\newblock \showarticletitle{Valgrind: a framework for heavyweight dynamic
  binary instrumentation}.
\newblock \bibinfo{journal}{\emph{Proceedings of the 28th ACM SIGPLAN
  Conference on Programming Language Design and Implementation}}
  (\bibinfo{year}{2007}), \bibinfo{pages}{89--100}.
\newblock
\urldef\tempurl%
\url{https://doi.org/10.1145/1250734.1250746}
\showDOI{\tempurl}


\bibitem[\protect\citeauthoryear{Ngyuen}{Ngyuen}{2020}]%
        {unicorn2github2020}
\bibfield{author}{\bibinfo{person}{Anh~Quynh Ngyuen}.}
  \bibinfo{year}{2020}\natexlab{}.
\newblock \bibinfo{title}{Unicorn 2 - looking for sponsors}.
\newblock
  \bibinfo{howpublished}{\url{https://github.com/unicorn-engine/unicorn/issues/1217}}.
\newblock


\bibitem[\protect\citeauthoryear{Ngyuen and Dang}{Ngyuen and Dang}{2015}]%
        {unicorn2015}
\bibfield{author}{\bibinfo{person}{Anh~Quynh Ngyuen} {and}
  \bibinfo{person}{Hoang~Vu Dang}.} \bibinfo{year}{2015}\natexlab{}.
\newblock \bibinfo{title}{Unicorn: Next Generation CPU Emulator Framework}.
\newblock
  \bibinfo{howpublished}{\url{http://www.unicorn-engine.org/BHUSA2015-unicorn.pdf}}.
\newblock


\bibitem[\protect\citeauthoryear{{\"O}sterlund, Razavi, Bos, and
  Giuffrida}{{\"O}sterlund et~al\mbox{.}}{2020}]%
        {parmesan2020}
\bibfield{author}{\bibinfo{person}{Sebastian {\"O}sterlund},
  \bibinfo{person}{Kaveh Razavi}, \bibinfo{person}{Herbert Bos}, {and}
  \bibinfo{person}{Cristiano Giuffrida}.} \bibinfo{year}{2020}\natexlab{}.
\newblock \showarticletitle{ParmeSan: Sanitizer-guided Greybox Fuzzing}. In
  \bibinfo{booktitle}{\emph{29th {USENIX} Security Symposium}}
  \emph{(\bibinfo{series}{USENIX Security '20})}. \bibinfo{pages}{2289--2306}.
\newblock
\urldef\tempurl%
\url{https://doi.org/10.5555/3489212.3489341}
\showDOI{\tempurl}


\bibitem[\protect\citeauthoryear{Poeplau and Francillon}{Poeplau and
  Francillon}{2021}]%
        {symqemu2021}
\bibfield{author}{\bibinfo{person}{Sebastian Poeplau} {and}
  \bibinfo{person}{Aurélien Francillon}.} \bibinfo{year}{2021}\natexlab{}.
\newblock \showarticletitle{{SymQEMU}: Compilation-based symbolic execution for
  binaries}. In \bibinfo{booktitle}{\emph{Network and Distributed System
  Security Symposium}} \emph{(\bibinfo{series}{NDSS '22})}.
\newblock
\urldef\tempurl%
\url{https://doi.org/10.14722/ndss.2021.23118}
\showDOI{\tempurl}


\bibitem[\protect\citeauthoryear{Scharnowski, Bars, Schloegel, Gustafson,
  Muench, Vigna, Kruegel, Holz, and Abbasi}{Scharnowski et~al\mbox{.}}{2022}]%
        {fuzzware2022}
\bibfield{author}{\bibinfo{person}{Tobias Scharnowski}, \bibinfo{person}{Nils
  Bars}, \bibinfo{person}{Moritz Schloegel}, \bibinfo{person}{Eric Gustafson},
  \bibinfo{person}{Marius Muench}, \bibinfo{person}{Giovanni Vigna},
  \bibinfo{person}{Christopher Kruegel}, \bibinfo{person}{Thorsten Holz}, {and}
  \bibinfo{person}{Ali Abbasi}.} \bibinfo{year}{2022}\natexlab{}.
\newblock \showarticletitle{Fuzzware: Using Precise {MMIO} Modeling for
  Effective Firmware Fuzzing}. In \bibinfo{booktitle}{\emph{31st {USENIX}
  Security Symposium}} \emph{(\bibinfo{series}{USENIX Security '22})}.
  \bibinfo{publisher}{USENIX Association}, \bibinfo{pages}{1239--1256}.
\newblock


\bibitem[\protect\citeauthoryear{Schumilo, Aschermann, Abbasi, W{\"o}rner, and
  Holz}{Schumilo et~al\mbox{.}}{2021}]%
        {nyx2020}
\bibfield{author}{\bibinfo{person}{Sergej Schumilo}, \bibinfo{person}{Cornelius
  Aschermann}, \bibinfo{person}{Ali Abbasi}, \bibinfo{person}{Simon
  W{\"o}rner}, {and} \bibinfo{person}{Thorsten Holz}.}
  \bibinfo{year}{2021}\natexlab{}.
\newblock \showarticletitle{Nyx: Greybox Hypervisor Fuzzing using Fast
  Snapshots and Affine Types}. In \bibinfo{booktitle}{\emph{30th {USENIX}
  Security Symposium}} \emph{(\bibinfo{series}{USENIX Security '21})}.
  \bibinfo{pages}{2597--2614}.
\newblock


\bibitem[\protect\citeauthoryear{Schumilo, Aschermann, Gawlik, Schinzel, and
  Holz}{Schumilo et~al\mbox{.}}{2017}]%
        {kafl2017}
\bibfield{author}{\bibinfo{person}{Sergej Schumilo}, \bibinfo{person}{Cornelius
  Aschermann}, \bibinfo{person}{Robert Gawlik}, \bibinfo{person}{Sebastian
  Schinzel}, {and} \bibinfo{person}{Thorsten Holz}.}
  \bibinfo{year}{2017}\natexlab{}.
\newblock \showarticletitle{kAFL: Hardware-assisted feedback fuzzing for {OS}
  kernels}. In \bibinfo{booktitle}{\emph{26th {USENIX} Security Symposium}}
  \emph{(\bibinfo{series}{USENIX Security '17})}. \bibinfo{pages}{167--182}.
\newblock
\urldef\tempurl%
\url{https://doi.org/10.5555/3241189.3241204}
\showDOI{\tempurl}


\bibitem[\protect\citeauthoryear{Schumilo, Aschermann, Jemmett, Abbasi, and
  Holz}{Schumilo et~al\mbox{.}}{2022}]%
        {nyxnet2022}
\bibfield{author}{\bibinfo{person}{Sergej Schumilo}, \bibinfo{person}{Cornelius
  Aschermann}, \bibinfo{person}{Andrea Jemmett}, \bibinfo{person}{Ali Abbasi},
  {and} \bibinfo{person}{Thorsten Holz}.} \bibinfo{year}{2022}\natexlab{}.
\newblock \showarticletitle{Nyx-Net: Network Fuzzing with Incremental
  Snapshots}. In \bibinfo{booktitle}{\emph{Proceedings of the 17th European
  Conference on Computer Systems}} \emph{(\bibinfo{series}{EuroSys '22})}.
  \bibinfo{pages}{166--180}.
\newblock
\urldef\tempurl%
\url{https://doi.org/10.1145/3492321.3519591}
\showDOI{\tempurl}


\bibitem[\protect\citeauthoryear{seal9055}{seal9055}{2022}]%
        {sfuzz2022}
\bibfield{author}{\bibinfo{person}{seal9055}.} \bibinfo{year}{2022}\natexlab{}.
\newblock \bibinfo{title}{SFUZZ: High Performance Coverage-guided Greybox
  Fuzzer with Custom JIT Engine}.
\newblock
  \bibinfo{howpublished}{\url{https://seal9055.com/blog/fuzzing/sfuzz}}.
\newblock


\bibitem[\protect\citeauthoryear{Smaragdakis and Bravenboer}{Smaragdakis and
  Bravenboer}{2010}]%
        {usingdatalog2010}
\bibfield{author}{\bibinfo{person}{Yannis Smaragdakis} {and}
  \bibinfo{person}{Martin Bravenboer}.} \bibinfo{year}{2010}\natexlab{}.
\newblock \showarticletitle{Using Datalog for fast and easy program analysis}.
  In \bibinfo{booktitle}{\emph{Datalog Reloaded - First International
  Workshop}} \emph{(\bibinfo{series}{Lecture Notes in Computer Science})}.
  \bibinfo{pages}{245--251}.
\newblock
\urldef\tempurl%
\url{https://doi.org/10.1007/978-3-642-24206-9_14}
\showDOI{\tempurl}


\bibitem[\protect\citeauthoryear{Song, Brumley, Yin, Caballero, Jager, Kang,
  Liang, Newsome, Poosankam, and Saxena}{Song et~al\mbox{.}}{2008}]%
        {bitblaze2008}
\bibfield{author}{\bibinfo{person}{Dawn Song}, \bibinfo{person}{David Brumley},
  \bibinfo{person}{Heng Yin}, \bibinfo{person}{Juan Caballero},
  \bibinfo{person}{Ivan Jager}, \bibinfo{person}{Min~Gyung Kang},
  \bibinfo{person}{Zhenkai Liang}, \bibinfo{person}{James Newsome},
  \bibinfo{person}{Pongsin Poosankam}, {and} \bibinfo{person}{Prateek Saxena}.}
  \bibinfo{year}{2008}\natexlab{}.
\newblock \showarticletitle{BitBlaze: A new approach to computer security via
  binary analysis}. In \bibinfo{booktitle}{\emph{International Conference on
  Information Systems Security}} \emph{(\bibinfo{series}{ICISS '08})}.
  \bibinfo{pages}{1--25}.
\newblock
\urldef\tempurl%
\url{https://doi.org/10.1007/978-3-540-89862-7_1}
\showDOI{\tempurl}


\bibitem[\protect\citeauthoryear{Tang and Li}{Tang and Li}{2016}]%
        {triforceafl2016}
\bibfield{author}{\bibinfo{person}{Jack Tang} {and} \bibinfo{person}{Moony
  Li}.} \bibinfo{year}{2016}\natexlab{}.
\newblock \showarticletitle{Project Triforce: Run {AFL} on Everything!}
\newblock \bibinfo{journal}{\emph{Black Hat Europe}} (\bibinfo{year}{2016}).
\newblock


\bibitem[\protect\citeauthoryear{{The National Security Agency (NSA)}}{{The
  National Security Agency (NSA)}}{2019}]%
        {ghidra2019}
\bibfield{author}{\bibinfo{person}{{The National Security Agency (NSA)}}.}
  \bibinfo{year}{2019}\natexlab{}.
\newblock \bibinfo{title}{Ghidra: Software Reverse Engineering Framework}.
\newblock \bibinfo{howpublished}{\url{https://ghidra-sre.org/}}.
\newblock


\bibitem[\protect\citeauthoryear{Travis}{Travis}{2021}]%
        {goodwatch2021}
\bibfield{author}{\bibinfo{person}{Goodspeed Travis}.}
  \bibinfo{year}{2021}\natexlab{}.
\newblock \bibinfo{title}{Goodwatch}.
\newblock
  \bibinfo{howpublished}{\url{https://github.com/travisgoodspeed/goodwatch}}.
\newblock


\bibitem[\protect\citeauthoryear{Wang, Duan, Song, Yin, and Song}{Wang
  et~al\mbox{.}}{2019}]%
        {aflsensitive2019}
\bibfield{author}{\bibinfo{person}{Jinghan Wang}, \bibinfo{person}{Yue Duan},
  \bibinfo{person}{Wei Song}, \bibinfo{person}{Heng Yin}, {and}
  \bibinfo{person}{Chengyu Song}.} \bibinfo{year}{2019}\natexlab{}.
\newblock \showarticletitle{Be sensitive and collaborative: Analyzing impact of
  coverage metrics in greybox fuzzing}. In \bibinfo{booktitle}{\emph{22nd
  International Symposium on Research in Attacks, Intrusions and Defenses}}
  \emph{(\bibinfo{series}{RAID '19})}. \bibinfo{pages}{1--15}.
\newblock


\bibitem[\protect\citeauthoryear{Yan and McCamant}{Yan and McCamant}{2018}]%
        {fastpokeemu2018}
\bibfield{author}{\bibinfo{person}{Qiuchen Yan} {and} \bibinfo{person}{Stephen
  McCamant}.} \bibinfo{year}{2018}\natexlab{}.
\newblock \showarticletitle{Fast PokeEMU: Scaling Generated Instruction Tests
  Using Aggregation and State Chaining}. In
  \bibinfo{booktitle}{\emph{Proceedings of the 14th ACM SIGPLAN/SIGOPS
  International Conference on Virtual Execution Environments}}
  \emph{(\bibinfo{series}{VEE '18})}. \bibinfo{pages}{71--83}.
\newblock
\urldef\tempurl%
\url{https://doi.org/10.1145/3186411.3186417}
\showDOI{\tempurl}


\bibitem[\protect\citeauthoryear{Zalewski}{Zalewski}{2010}]%
        {afl2010}
\bibfield{author}{\bibinfo{person}{Michal Zalewski}.}
  \bibinfo{year}{2010}\natexlab{}.
\newblock \bibinfo{title}{American Fuzzy Lop: a security-oriented fuzzer}.
\newblock \bibinfo{howpublished}{\url{https://github.com/google/AFL}}.
\newblock
\urldef\tempurl%
\url{https://lcamtuf.coredump.cx/afl/}
\showURL{%
\tempurl}


\bibitem[\protect\citeauthoryear{Zheng, Davanian, Yin, Song, Zhu, and
  Sun}{Zheng et~al\mbox{.}}{2019}]%
        {firmafl2019}
\bibfield{author}{\bibinfo{person}{Yaowen Zheng}, \bibinfo{person}{Ali
  Davanian}, \bibinfo{person}{Heng Yin}, \bibinfo{person}{Chengyu Song},
  \bibinfo{person}{Hongsong Zhu}, {and} \bibinfo{person}{Limin Sun}.}
  \bibinfo{year}{2019}\natexlab{}.
\newblock \showarticletitle{{FIRM-AFL}: High Throughput Greybox Fuzzing of IoT
  Firmware via Augmented Process Emulation}. In \bibinfo{booktitle}{\emph{28th
  {USENIX} Security Symposium}} \emph{(\bibinfo{series}{USENIX Security '19})}.
  \bibinfo{pages}{1099--1114}.
\newblock
\urldef\tempurl%
\url{https://doi.org/10.5555/3361338.3361415}
\showDOI{\tempurl}


\bibitem[\protect\citeauthoryear{Zheng, Li, Zhang, Zhu, Liu, and Sun}{Zheng
  et~al\mbox{.}}{2022}]%
        {equafl2022}
\bibfield{author}{\bibinfo{person}{Yaowen Zheng}, \bibinfo{person}{Yuekang Li},
  \bibinfo{person}{Cen Zhang}, \bibinfo{person}{Hongsong Zhu},
  \bibinfo{person}{Yang Liu}, {and} \bibinfo{person}{Limin Sun}.}
  \bibinfo{year}{2022}\natexlab{}.
\newblock \showarticletitle{Efficient greybox fuzzing of applications in
  Linux-based IoT devices via enhanced user-mode emulation}. In
  \bibinfo{booktitle}{\emph{Proceedings of the 31st ACM SIGSOFT International
  Symposium on Software Testing and Analysis}} \emph{(\bibinfo{series}{ISSTA
  '22})}. \bibinfo{pages}{417--428}.
\newblock
\urldef\tempurl%
\url{https://doi.org/10.1145/3533767.3534414}
\showDOI{\tempurl}


\bibitem[\protect\citeauthoryear{Zhou, Guan, Liu, and Zhang}{Zhou
  et~al\mbox{.}}{2021}]%
        {uemu2021}
\bibfield{author}{\bibinfo{person}{Wei Zhou}, \bibinfo{person}{Le Guan},
  \bibinfo{person}{Peng Liu}, {and} \bibinfo{person}{Yuqing Zhang}.}
  \bibinfo{year}{2021}\natexlab{}.
\newblock \showarticletitle{Automatic Firmware Emulation through
  Invalidity-guided Knowledge Inference}. In \bibinfo{booktitle}{\emph{30th
  {USENIX} Security Symposium}} \emph{(\bibinfo{series}{USENIX Security '21})}.
  \bibinfo{pages}{2007--2024}.
\newblock
\showISBNx{978-1-939133-24-3}


\end{thebibliography}

\end{document}